%% file: emnlp2023.tex
\pdfoutput=1

\documentclass[11pt]{article}

\usepackage{EMNLP2023}

\usepackage{times}
\usepackage{latexsym}

\usepackage[T1]{fontenc}

\usepackage[utf8]{inputenc}

\usepackage{microtype}

\usepackage{inconsolata}

\usepackage{xspace}
\usepackage{booktabs}
\usepackage{caption}
\usepackage{subcaption}
\usepackage{graphicx}
\usepackage{float}
\usepackage[normalem]{ulem}
\usepackage{multirow}
\usepackage{amsmath}
\usepackage{enumitem}
\usepackage{balance}
\usepackage{algorithm}
\usepackage[noend]{algorithmic}
\usepackage{mathtools}
\usepackage[mathscr]{euscript}

\usepackage{mkolar_definitions}
\usepackage{balance}

\input{notation}

\newcount\Comments  
\definecolor{purple}{rgb}{0.5,0,1}
\definecolor{teal}{rgb}{0.33,0.65,0.55}
\definecolor{green}{rgb}{0.1,0.65,0.1}
\newcommand{\kibitz}[2]{\ifnum\Comments=1\textcolor{#1}{#2}\fi}

\title{Efficient $k$-NN Search with Cross-Encoders \\ using Adaptive Multi-Round CUR Decomposition}

\author{Nishant Yadav$^\dag$, Nicholas Monath$^{\diamondsuit}$,   \textbf{Manzil Zaheer}$^{\diamondsuit}$,  and \textbf{Andrew McCallum}$^\dag$ \\
$^\dag$ University of Massachusetts Amherst \quad $^{\diamondsuit}$Google Research \\
\texttt{\{nishantyadav, mccallum\}@cs.umass.edu} \quad \texttt{\{nmonath,manzilzaheer\}@google.com}
}

\begin{document}
\maketitle

\input{tex/00_abstract.tex}
\input{tex/01_intro.tex}

\input{tex/02_model.tex}

\input{tex/03_experiments.tex}

\input{tex/05_conclusion.tex}

\input{tex/06_limitations_ethics.tex}
\input{tex/07_ack.tex}

\bibliography{emnlp2023}
\bibliographystyle{acl_natbib}

\appendix

\input{tex/08_appendix.tex}

\end{document}

%% file: notation.tex
\newcommand{\tfidf}{\textsc{tf-idf}\xspace}

\newcommand{\zeshel}{\textsc{ZeShEL}\xspace}
\newcommand{\beir}{\textsc{BeIR}\xspace}
\newcommand{\muver}{\textsc{Muver}\xspace}

\newcommand{\eCrossenc}{\textsc{[emb]-CE}\xspace}

\newcommand{\yugioh}{\texttt{YuGiOh}\xspace}
\newcommand{\starTrek}{\texttt{StarTrek}\xspace}

\newcommand{\military}{\texttt{Military}\xspace}

\newcommand{\scidocs}{\texttt{SciDocs}\xspace}
\newcommand{\hotpotqa}{\texttt{HotpotQA}\xspace}

\newcommand{\query}{q\xspace}
\newcommand{\querySpace}{\mathcal{Q}\xspace}
\newcommand{\testQuery}{q_\textsubscript{{\texttt{test}}}\xspace}
\newcommand{\dataItem}{i\xspace}
\newcommand{\itemSpace}{\mathcal{I}\xspace}
\newcommand{\anchorItems}{\mathcal{I}_{\texttt{anc}}\xspace}
\newcommand{\anchorQueries}{\queryTrainData}

\newcommand{\rowMatrixTrain}{R_\text{anc}\xspace}

\newcommand{\colMatrixTest}{C_\text{test}\xspace}

\newcommand{\itemRep}{E^{\itemSpace}\xspace}

\newcommand{\nAnchorQueries}{k_q\xspace}
\newcommand{\nAnchorItems}{k_i\xspace}

\newcommand{\nItems}{\lvert \itemSpace \rvert\xspace}

\newcommand{\queryEmbed}[1][q]{e_{#1}\xspace}
\newcommand{\cost}{\mathcal{C}}

\newcommand{\model}[1]{f_{#1}\xspace}

\newcommand{\bert}{\textsc{Bert}\xspace}

\newcommand{\fixedDualEncoder}{\textsc{DE\textsubscript{base}}\xspace}
\newcommand{\finetuneDualEncoder}{$\textsc{DE}^{\textsuperscript{\textsc{ce}}}_{\textsc{base}}$\xspace}

\newcommand{\scratchDualEncoder}{$\textsc{DE}^{\textsuperscript{\textsc{ce}}}_{\textsc{bert}}$\xspace}

\newcommand{\annCUR}{\textsc{annCUR}\xspace}
\newcommand{\annCURwFixedDE}{\textsc{annCUR}\textsubscript{\texttt{DE\textsubscript{BASE}}\xspace}}
\newcommand{\annCURwTFIDF}{\textsc{annCUR}\textsubscript{\texttt{TF-IDF}\xspace}}

\newcommand{\maskedItems}{\mathcal{I}_{\texttt{mask}}\xspace}
\newcommand{\sampledItems}{\mathcal{I}_{\texttt{select}}\xspace}

\newcommand{\adaCUR}{\textsc{adaCUR}\xspace}
\newcommand{\adaCURwFixedDE}{\textsc{adaCUR}\textsubscript{\texttt{DE\textsubscript{BASE}}}\xspace}
\newcommand{\adaCURwTFIDF}{\textsc{adaCUR}\textsubscript{\texttt{TF-IDF}\xspace}}

\newcommand{\queryTestData}{\mathcal{Q}_\text{test}}
\newcommand{\queryTrainData}{\mathcal{Q}_\text{train}}
\newcommand{\queryTrainSize}{\lvert \queryTrainData \rvert}

\newcommand{\ceBudget}{\mathcal{B}_\text{CE}}
\newcommand{\nRounds}{N_{\mathcal{R}}}
\newcommand{\roundIter}{j}
\newcommand{\nAnchorItemsPerRound}{k_s}
\newcommand{\sampleMethod}{\mathcal{A}}

\newcommand{\topkSample}{\texttt{TopK}\xspace}
\newcommand{\softmaxSample}{\texttt{SoftMax}\xspace}

\newcommand{\epsTopkGTSample}[2]{\topkSample^\mathcal{O}_{#1,#2}}
\newcommand{\epsSoftmaxGTSample}[2]{\softmaxSample^\mathcal{O}_{#1,#2}}
\newcommand{\Kmask}{k_m}

%% file: tex/00_abstract.tex
\begin{abstract}
Cross-encoder models, which jointly encode and score a query-item pair, are prohibitively expensive for direct $k$-nearest neighbor ($k$-NN) search.
Consequently, $k$-NN search typically employs a fast approximate retrieval (e.g. using BM25 or dual-encoder vectors), followed by reranking with a cross-encoder; however, the retrieval approximation often has detrimental recall regret.
This problem is tackled by \annCUR~\cite{yadav2022efficient}, a recent work that employs a cross-encoder only, making search efficient using a relatively small number of anchor items, and a CUR matrix factorization.
While \annCUR's one-time selection of anchors tends to approximate the cross-encoder distances on average,
doing so forfeits the capacity to accurately estimate distances to items near the query, leading to 
regret in the crucial end-task: recall of top-$k$ items.
In this paper, we propose \adaCUR, a method that adaptively, iteratively, and efficiently 
minimizes the approximation error for the practically important top-$k$ neighbors.
It does so by iteratively performing $k$-NN search using the anchors available so far, then adding these retrieved nearest neighbors to the anchor set for the next round.
Empirically, on multiple datasets, in comparison to previous traditional and state-of-the-art methods such as \annCUR and dual-encoder-based retrieve-and-rerank, our proposed approach \adaCUR consistently reduces recall error---by up to 70\% on the important $k=1$ setting---while using no more compute than its competitors.
\end{abstract}

%% file: tex/01_intro.tex
\section{Introduction}
$k$-nearest neighbor~($k$-NN) search is a core sub-routine 
of a variety of tasks in NLP such as entity linking~\cite{wu-etal-2020-scalable},
passage retrieval for QA~\cite{karpukhin2020dense}, and more generally,
in retrieval-augmented machine learning models \cite{guu2020retrieval,izacard2022few}.
For many of these applications, the state-of-the-art similarity function
is a cross-encoder that directly outputs a scalar similarity score after
jointly encoding a given query-item pair.
However, computing a \emph{single} query-item score using a cross-encoder 
requires a forward pass of the model which can be
computationally expensive as cross-encoders are typically
parameterized using deep neural models such as transformers~\cite{vaswani2017attention}. 
For this reason, $k$-NN search with a cross-encoder typically
involves retrieving candidate items using additional models
such as dual-encoders or BM25~\cite{robertson2009probabilistic}, followed by re-ranking items 
using the cross-encoder~\cite{logeswaran-etal-2019-zero,  zhang-stratos-2021-understanding, qu-etal-2021-rocketqa}.
However, the accuracy of such retrieve-and-rerank 
approaches is upperbound by the recall of first-stage
retrieval and may require computationally expensive
distillation-based training of dual-encoders to improve recall.

\input{figs/plots_random_anchors.tex}

Recent work by~\citet{yadav2022efficient} proposed \annCUR, a CUR factorization~\cite{mahoney2009cur} 
based method, that approximates cross-encoder using dot-product of latent
query and item embeddings and performs $k$-NN retrieval using approximate
scores followed by optionally re-ranking retrieved items using exact cross-encoder scores. 
The latent item embeddings are computed 
by comparing each item against a set of anchor queries,
and the latent query embedding is computed using 
the query's cross-encoder scores against a fixed set of anchor items.
As shown in Figure~\ref{fig:score_dist_example_random_sampling},
when \annCUR selects the anchor items uniformly at 
random~(Fig~\ref{fig:score_dist_example_random_sampling_50}),
it incurs higher approximation error on the top-scoring items 
than the rest of the items, resulting in poor $k$-NN recall,
and including some $k$-NN items as part of anchor 
items~(Fig.~\ref{fig:score_dist_example_topk_sampling_50}) 
can significantly improve approximation error for top-scoring items.

In this work, we propose \adaCUR, a search strategy that
improves $k$-NN search recall by improving the approximation
of top-scoring items.
\adaCUR performs retrieval over multiple rounds,
retrieving the first batch of items either uniformly at random
or using heuristic or auxiliary models such as dual-encoder or BM25
to get a first coarse approximation of item scores for the test query.
In subsequent rounds, it alternates between 
a) performing retrieval using approximate scores 
and scoring retrieved items using cross-encoder and 
b) using all retrieved items as anchor items to improve the approximation and hence retrieval of relevant items in the subsequent rounds.
Our proposed approach provides significant improvements 
in $k$-NN search recall  over \annCUR and dual-encoder based 
retrieve-and-rerank approaches
when performing $k$-NN search with cross-encoder models
trained for the task of entity linking and 
information retrieval.

%% file: figs/plots_random_anchors.tex
\begin{figure}[!t]
    \centering
    \begin{subfigure}[b]{0.4\textwidth}
         \includegraphics[width=\textwidth]{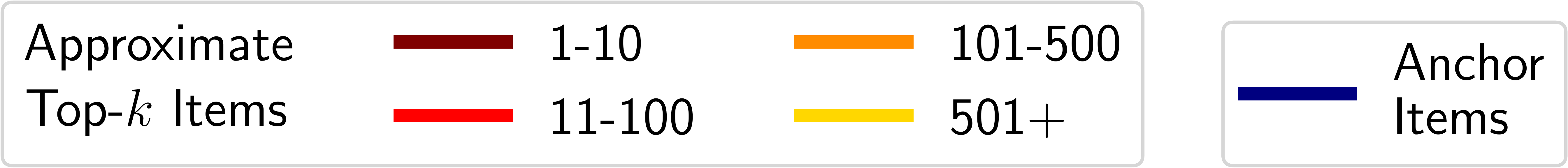}
        \phantomcaption{} 
    \end{subfigure}
    \addtocounter{subfigure}{-1}
    \begin{subfigure}[b]{0.23\textwidth}
         \includegraphics[width=\textwidth]{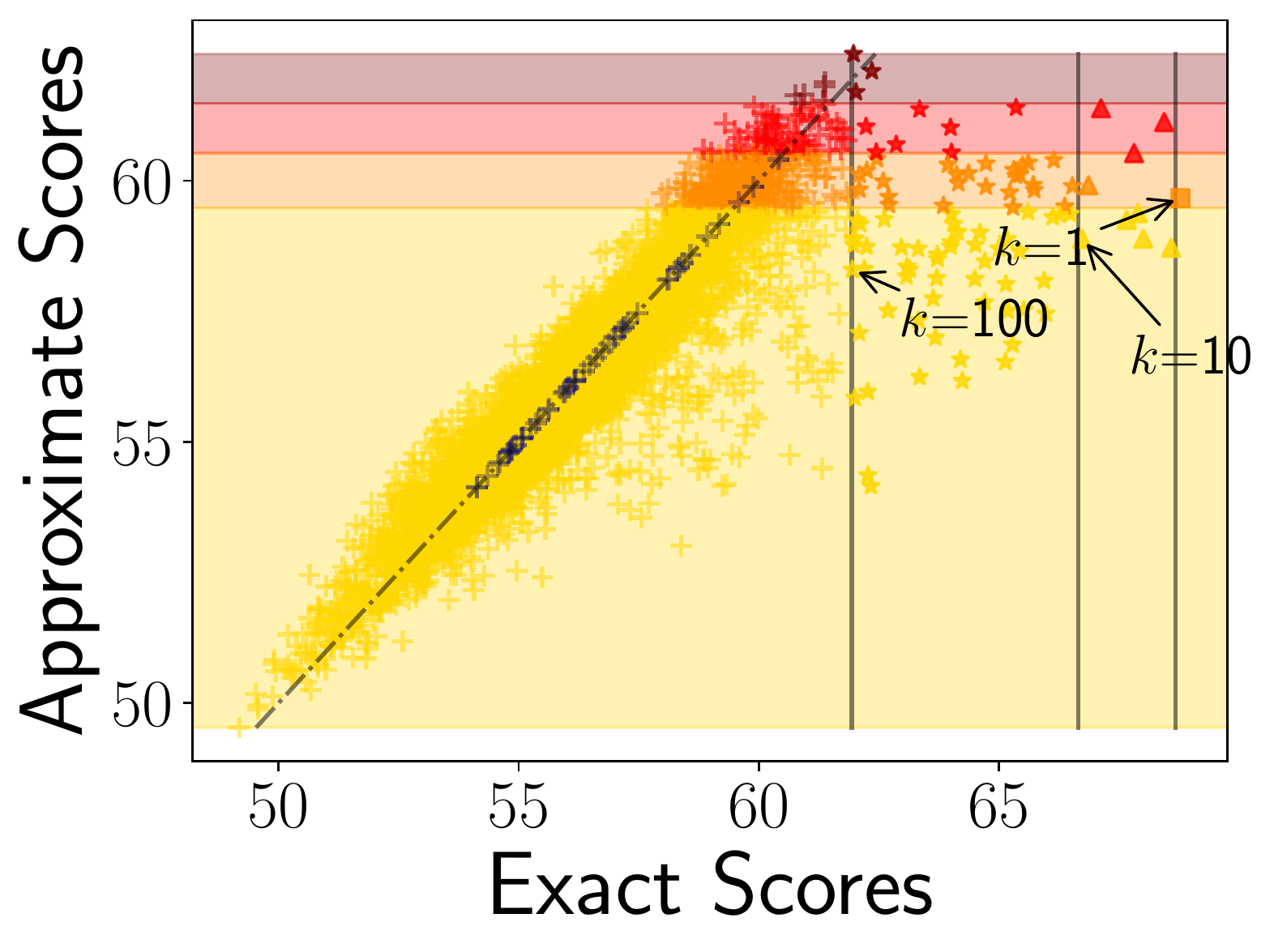}
        \caption{Uniformly at random}  
        \label{fig:score_dist_example_random_sampling_50}
    \end{subfigure}
    \hfill 
    \begin{subfigure}[b]{0.23\textwidth}
         \includegraphics[width=\textwidth]{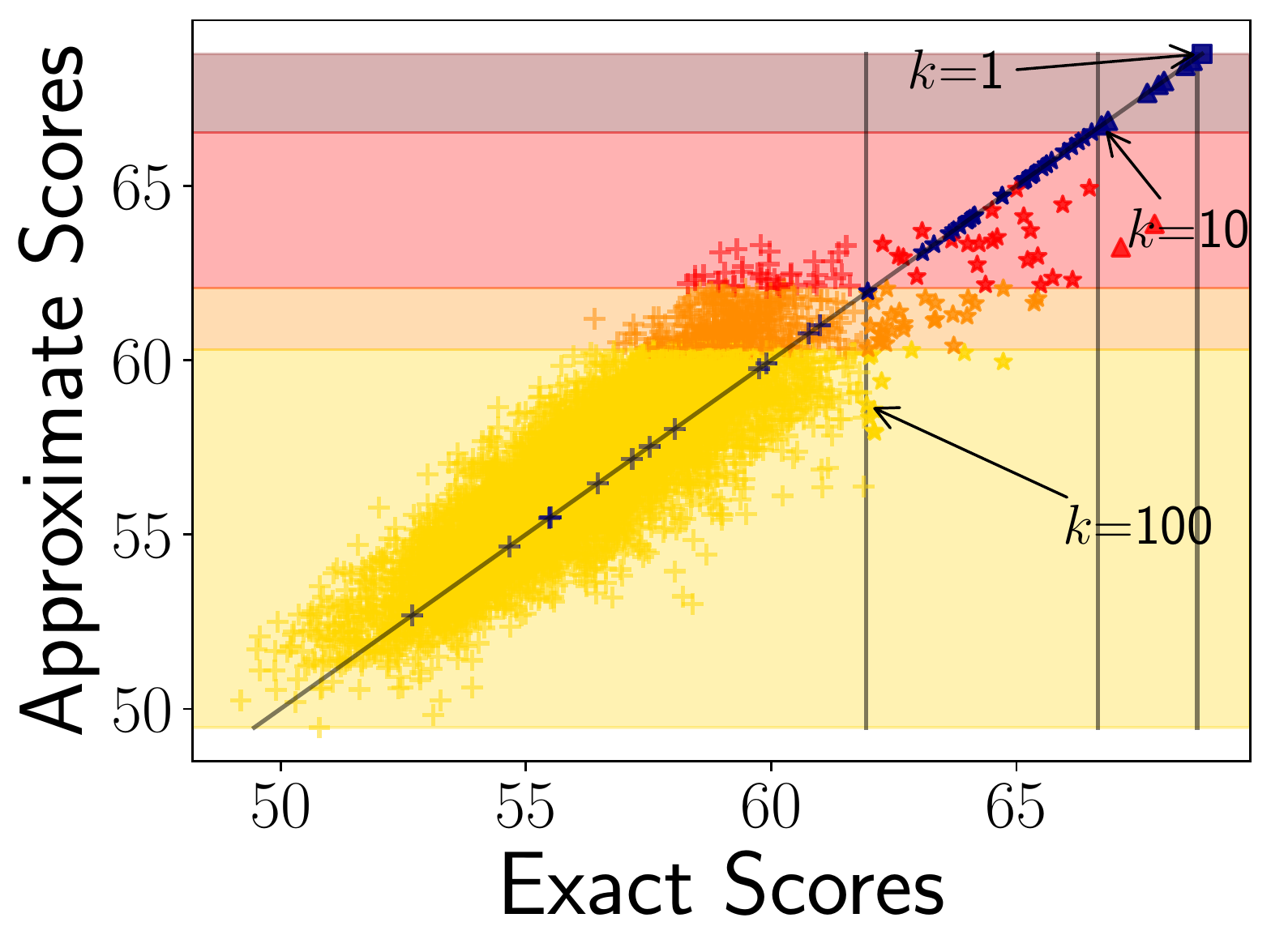}
        \caption{Biased sampling}  
        \label{fig:score_dist_example_topk_sampling_50}
    \end{subfigure}
    \caption{
    Exact versus approximate cross-encoder scores (computed using \annCUR) of all items for 
    a test-query in domain=\yugioh.
    \annCUR incurs high approximation error on $k$-NN items wrt exact scores
    when using 50 anchor items sampled uniformly at random~(Fig.~\ref{fig:score_dist_example_random_sampling_50}).
    In contrast, sampling 50 anchor items with probability proportional
    to exact cross-encoder scores~(Fig.~\ref{fig:score_dist_example_topk_sampling_50})
    significantly improves approximation of top-scoring items.
    }
    \vspace{-0.6cm}
    \label{fig:score_dist_example_random_sampling}
\end{figure}

%% file: tex/02_model.tex
\section{Proposed Method: \adaCUR}
\label{subsec:proposed_method}

\paragraph{Task}
Given a scoring function
$\model{\theta} \colon \querySpace \times \itemSpace \rightarrow \RR $ 
that maps a query-item pair to a scalar score, and a
query $\query \in \querySpace$, the $k$-nearest neighbor search 
task is to retrieve top-$k$ scoring items 
from a fixed item set $\itemSpace$ 
according to the given scoring function $\model{\theta}$.

\subsection{\adaCUR: Offline Indexing of Items}
\label{subsec:adacur_indexing}
The indexing step of \adaCUR involves 
using the cross-encoder model~$(\model{\theta})$ to score each item against 
a fixed set of  $\nAnchorQueries$ anchor/train queries ($\anchorQueries$), to get score matrix $\rowMatrixTrain$
\begin{align*}
   \rowMatrixTrain(q,i) &= \model{\theta}(q, i), \hfill \text{ }  \forall (q,i) \in \anchorQueries \times \itemSpace
   \vspace{-0.7cm}
\end{align*}
\noindent Each column of $\itemRep \coloneqq \rowMatrixTrain \in \RR^{\nAnchorQueries \times \nItems}$ 
corresponds to a $\nAnchorQueries$-dimensional latent item embedding.

\input{tables_and_algos/algo_inference.tex}

\input{tables_and_algos/algo_sample_anchors.tex}
\subsection{\adaCUR: Test-time inference}
\label{subsec:adacur_inference}
The baseline method \annCUR computes the
latent test-query embedding~$\queryEmbed[\testQuery]$ 
using $C_{\testQuery}$, a $\lvert \anchorItems \rvert$-dimensional
vector containing 
cross-encoder scores of $\testQuery$ with a set of anchor items ($\anchorItems$) as:
\begin{align*}
    C_{\testQuery} &= [  \model{\theta}(\testQuery, \dataItem) ]_{\dataItem \in \anchorItems} \\
    \queryEmbed[\testQuery] &= C_{\testQuery} \times U 
\end{align*}
where $U \in \RR^{ \lvert \anchorItems \rvert \times \lvert \anchorQueries \rvert  }$ is the pseudo-inverse of 
$\rowMatrixTrain[\anchorItems]$, the subset of columns of $\rowMatrixTrain$ corresponding to (anchor) items $\anchorItems$.
Finally, \annCUR approximates the score for a 
query-item pair ($\testQuery, \dataItem$) using
dot-product of the query embedding $\queryEmbed[\testQuery]$ and item embedding $\itemRep[:,\dataItem]$ as
\vspace{-0.2cm}
\begin{equation*}    
    \hat{\model{\theta}}(\testQuery,\dataItem) = \queryEmbed[\testQuery]^{\top} \itemRep[:,\dataItem]
    \vspace{-0.2cm}
\end{equation*}

The main bottleneck at test-time inference is the number 
of items scored using the cross-encoder for the given test-query.
For a given budget of cross-encoder calls, \annCUR splits the budget~($\ceBudget$) into two
parts -- it uses $\nAnchorItems$ cross-encoder calls to compare against anchor items 
(chosen uniformly at random or using heuristic methods)
and retrieves $\ceBudget - \nAnchorItems$ items using approximate scores
and re-ranks them using exact cross-encoder scores.

In contrast, our proposed approach \adaCUR uses the cross-encoder call budget to adaptively retrieve and score items
over $\nRounds$ rounds, re-purposing the retrieved items as anchor items
as shown in Algorithm~\ref{alg:incremental_cur}.
\adaCUR begins by sampling the first batch of $\nAnchorItemsPerRound=\ceBudget/\nRounds$ (anchor) items
uniformly at random.
The first batch of items can also be selected using 
baseline retrieval methods such as BM25 and dual-encoders.
In the $\roundIter\textsuperscript{th}$ round, the items retrieved
up to round $\roundIter-1$ are used as anchor items to revise 
the test-query embedding, which in turn is used to 
update the approximate scores~(line~\ref{alg_line:approx_score_cur_mat_mul} in Algorithm~\ref{alg:incremental_cur}).
Finally, the items selected so far are masked out and 
the next batch of $\nAnchorItemsPerRound$ items in round $\roundIter$ 
is retrieved using the updated approximate scores in the following two ways: 
\begin{itemize}[topsep=0pt,itemsep=0ex,partopsep=1ex,parsep=1ex]
    \item \topkSample : Greedily pick top-$\nAnchorItemsPerRound$ items 
    according to approximate scores.
    \item \softmaxSample : Convert approximate item scores into  probability using softmax and sample $\nAnchorItemsPerRound$ items without replacement.
\end{itemize}
Finally, \adaCUR returns top-$k$ items based on exact cross-encoder 
scores\footnote{Sorting retrieved items based on exact cross-encoder scores does not require
any additional cross-encoder calls as cross-encoder scores for these items have already been computed~(see line~\ref{alg_line:update_ce_scores} in Algorithm~\ref{alg:incremental_cur}).}
from the set of retrieved (anchor) items as approximate $k$-NN items.
We refer interested readers to Appendix~\ref{apndx_subsec:time_complexity} 
for a discussion on the time complexity of \adaCUR.

%% file: tables_and_algos/algo_inference.tex
\begin{algorithm}[!t]
\caption{\adaCUR $k$-NN Search}
\begin{algorithmic}[1]
\small
\STATE \textbf{Input:} $(\testQuery, \rowMatrixTrain, \nRounds, \ceBudget, \sampleMethod)$
\STATE $\testQuery$: Test query
\STATE $\rowMatrixTrain$: Matrix containing CE scores between $\anchorQueries$ and $\itemSpace$
\STATE $\ceBudget$: Total cross-encoder (CE) call budget.
\STATE $\sampleMethod$: Algorithm to use for selecting (anchor) items.
\STATE $\nRounds$: Number of iterative search rounds
\STATE \textbf{Output:} $\hat{S}$: Approximate scores of $\testQuery$ with all items,
$\anchorItems$: Retrieved (anchor) items with CE scores in $\colMatrixTest$.
\item[]
\STATE $\nAnchorItemsPerRound \gets \ceBudget/\nRounds$ \hfill $\rhd$ \textcolor{gray}{Num. of items to sample per round}
\STATE $\anchorItems \gets \textsc{Init}(\itemSpace, \nAnchorItemsPerRound)$ \hfill $\rhd$ \textcolor{gray}{Initial set of anchor items} \label{alg_line:init_incremental_cur}
\STATE $\colMatrixTest \gets [\model{\theta}(\testQuery, \dataItem)]_{\dataItem \in \anchorItems}$ \hfill $\rhd$ \textcolor{gray}{CE scores of $\anchorItems$ for $\testQuery$}

\FOR {$\roundIter \gets 2 \text{ to } \nRounds$}
    \STATE $U \gets \rowMatrixTrain[\anchorItems]^{\dagger}$  \hfill $\rhd$ \textcolor{gray}{$U \in \mathbb{R}^{ \lvert \anchorItems \rvert  \times \lvert \anchorQueries \rvert}$} \label{alg_line:approx_score_cur_inv}
    \STATE $\hat{S}^{(\roundIter)} \gets \colMatrixTest \times U \times \rowMatrixTrain$ \hfill $\rhd$ \textcolor{gray}{Update approx. scores} \label{alg_line:approx_score_cur_mat_mul}
    
    \STATE $\anchorItems^{(\roundIter)} \gets \textsc{SampleItems}(\sampleMethod, \nAnchorItemsPerRound, \anchorItems, \hat{S}^{(\roundIter)})$  
    \STATE $\anchorItems \gets \anchorItems \cup \anchorItems^{(\roundIter)}$
    \STATE $\colMatrixTest \gets \colMatrixTest \oplus [\model{\theta}(\testQuery, \dataItem)]_{\dataItem \in \anchorItems^{(\roundIter)}}$ \hfill $\rhd$ \textcolor{gray}{Update $\colMatrixTest$} \label{alg_line:update_ce_scores}
\ENDFOR

\STATE $U \gets \rowMatrixTrain[\anchorItems]^{\dagger}$  \hfill $\rhd$ \textcolor{gray}{$U \in \mathbb{R}^{ \lvert \anchorItems \rvert  \times \lvert \anchorQueries \rvert}$} 
\STATE $\hat{S} \gets \colMatrixTest \times U \times \rowMatrixTrain$  \hfill $\rhd$ \textcolor{gray}{Compute approx. scores}
    
\STATE \textbf{return} $\hat{S}, \anchorItems, \colMatrixTest$ 
\end{algorithmic}
\label{alg:incremental_cur}
\end{algorithm}

%% file: tables_and_algos/algo_sample_anchors.tex
\begin{algorithm}[!t]
\caption{\textsc{SampleItems}}
\begin{algorithmic}[1]
\footnotesize
\STATE \textbf{Input: $(\sampleMethod, \nAnchorItemsPerRound, \maskedItems, S)$} 
\STATE $\sampleMethod$: Algorithm for sampling items
\STATE $\nAnchorItemsPerRound$: Number of items to sample
\STATE $\maskedItems$ : Set of items to mask
\STATE $S$: (Approximate) Scores for all items
\STATE \textbf{Output:} $\sampledItems$ :  Set of sampled $\nAnchorItemsPerRound$ items
\item[]
\STATE $\bar{S} \gets \textsc{SoftMax}(S)$
\STATE $\bar{S}[\maskedItems] \gets 0$ \hfill $\rhd$ \textcolor{gray}{Mask items in $\maskedItems$} 
\IF {$\sampleMethod$ = \topkSample}
    \STATE $\sampledItems \gets $ \textsc{TopK}($\bar{S}, \nAnchorItemsPerRound$)
\ELSIF {$\sampleMethod$ = \softmaxSample}
    \STATE $\sampledItems \gets \nAnchorItemsPerRound$ items sampled using $\bar{S}$ 
\ELSIF {$\sampleMethod$ = \texttt{Random}}
    \STATE $\sampledItems \gets \nAnchorItemsPerRound$ items uniformly sampled from $\itemSpace \setminus \maskedItems $
\ENDIF
\STATE \textbf{return} $\sampledItems$
\end{algorithmic}
\label{alg:sample_anchors}
\end{algorithm}

%% file: tex/03_experiments.tex
\section{Experiments}
\label{sec:experiments}
In our experiments, we evaluate the proposed approach 
and baselines on the task
of finding $k$-nearest neighbor items as per a given
cross-encoder.  
We experiment with two cross-encoders --
one trained for the task of zero-shot entity linking,
and another trained on information 
retrieval datasets.

\input{figs/plots_rq_2.tex}

\paragraph{Experimental Setup} 
We run evaluation on domains \yugioh, \starTrek, and \military from \zeshel--a zero-shot
entity linking dataset~\cite{logeswaran-etal-2019-zero}, 
and domains \scidocs and \hotpotqa from \beir--a zero-shot
information retrieval benchmark~\cite{thakur2021beir}.
We use two cross-encoder models 
trained on labeled training data
from the corresponding benchmark and evaluate 
separately on each domain on the task
of finding $k$-NN cross-encoder items.
For each \zeshel domain, we randomly split the query set into a 
train/anchor set ($\queryTrainData$) and a test set ($\queryTestData$) while for \beir domains,
we use pseudo-queries released as part of the benchmark 
as train/anchor queries and evaluate on queries in the official test split.
We refer the reader to Table~\ref{tab:dataset_stats} for additional details.

\paragraph{Baselines}
We compare our proposed approach with the following baseline retrieval methods.

\noindent \textbf{Dual-Encoders~(DE)}: Query-item scores are computed using
dot-product of embeddings produced by a learned deep encoder model.
DE is used for initial retrieval followed by re-ranking using the cross-encoder.
We report results for \fixedDualEncoder, a dual-encoder trained 
on training domains in the corresponding dataset, 
and the following two dual-encoder models trained on the target domain 
via distillation using the cross-encoder.
\begin{itemize}[topsep=1pt,itemsep=0ex,partopsep=1ex,parsep=1ex]
    \item \scratchDualEncoder:  DE initialized with 
        \bert~\cite{devlin-etal-2019-bert} and trained \emph{only} on
        the target domain via distillation using the cross-encoder.
    \item \finetuneDualEncoder: \fixedDualEncoder model further fine-tuned  
        on the target domain via distillation. 
\end{itemize}
\noindent \textbf{\annCUR} : $k$-NN search method proposed by~\citet{yadav2022efficient} where anchor items are chosen uniformly at random. We additionally experiment
with $\annCURwFixedDE$ which uses top-scoring items retrieved using  \fixedDualEncoder as anchor items.

\paragraph{Evaluation Metric}
Following the precedent set by previous work~\cite{yadav2022efficient},
we evaluate all approaches using Top-$k$-Recall@$\ceBudget$
which is defined as the fraction of $k$-NN items retrieved under test-time cost budget $\ceBudget$
where the cost is defined as the number of cross-encoder calls made during 
inference. DE baselines will use the entire budget of $\ceBudget$ calls
for re-ranking retrieved items using exact cross-encoder scores, \annCUR
splits the budget between scoring anchor $\nAnchorItems$ items 
and using exact cross-encoder scores for re-ranking $\ceBudget - \nAnchorItems$ retrieved items,
and \adaCUR use the budget to adaptively search for $k$-NN items. 

For \adaCUR, unless noted otherwise, we use $\nRounds=5$ with 
$\topkSample$ method for retrieving items using approximate scores, 
and retrieve the first batch of items uniformly at random (\adaCUR) or 
using \fixedDualEncoder ($\adaCURwFixedDE$).
We refer readers to Appendix~\ref{apndx_sec:training_details} for implementation details and details
on training and parameterization of cross-encoder and dual-encoder models used in our experiments.

\subsection{Results}

\paragraph{\adaCUR versus baselines}
Figure~\ref{fig:rq_2_recall_at_same_cost_yugioh} shows 
Top-$k$-Recall for \adaCUR and baselines on domain=\yugioh.
\adaCUR, which uses adaptively retrieved items as anchor items 
over $\nRounds=5$ rounds, consistently outperforms \annCUR
which selects \emph{all} anchor items uniformly at random.
\adaCUR also outperforms strong DE baseline 
\scratchDualEncoder for all values of $k$ and 
outperforms \fixedDualEncoder \& \finetuneDualEncoder 
for large values of $k=10,100$.

\paragraph{Sampling anchor items using \fixedDualEncoder}
For $k=1, 10$, Top-$k$-Recall for both \annCUR and \adaCUR 
can be further improved by leveraging baseline retrieval models such as 
\fixedDualEncoder for retrieving all and 
the first batch of anchor items respectively instead 
of sampling them uniformly at random.
\adaCURwFixedDE always performs better
than \annCURwFixedDE which in turn performs better than re-ranking items 
retrieved using \fixedDualEncoder.
Thus, for a given cost budget~($\ceBudget$), even when a strong
baseline retrieval model such as \fixedDualEncoder is available, 
using the baseline retrieval model to select the first batch of
$\ceBudget/\nRounds$ items followed by using the proposed
approach to adaptively retrieve more items over remaining $\nRounds-1$ rounds 
performs better than merely re-ranking $\ceBudget$ top-scoring items
retrieved using \fixedDualEncoder. 
Finally, note that \adaCURwFixedDE outperforms all 
baselines including \finetuneDualEncoder which requires additional compute- and resource-intensive 
distillation-based fine-tuning of \fixedDualEncoder on the target domain.

We refer the reader to Appendix~\ref{apndx_sec:additional_results} for results
on other domains, training data size~($\queryTrainSize$) values, 
oracle-based sampling experiments~(\S{\ref{apndx_subsec:oracle_exp}}),
and additional results for multi-vector encoder~(\S{\ref{apndx_subsec:multi_vec_results}}) and \tfidf~(\S{\ref{apndx_subsec:tfidf_results}}) baselines.

\input{figs/plots_rq_3.tex}

\input{figs/plots_rq_4.tex}

\paragraph{Effect of number of rounds}
Figure~\ref{fig:rq_3_recall_at_same_cost_yugioh_vs_n_steps} 
shows Top-$k$-Recall and Figure~\ref{fig:rq_4_latency_breakdown_yugioh} shows total inference 
latency of \adaCUR on primary y-axis for various values of the test-time cross-encoder call budget~($\ceBudget$). 
The secondary y-axis in Figure~\ref{fig:rq_4_latency_breakdown_yugioh} 
shows the fraction of total inference time spent on each one of the 
three main steps in Algorithm~\ref{alg:incremental_cur} --  
(a) computing $U$ as pseudo-inverse of $\rowMatrixTrain[\anchorItems]$~(line~\ref{alg_line:approx_score_cur_inv}), 
(b) updating approximate scores for all items~(line~\ref{alg_line:approx_score_cur_mat_mul}), and
(c) computing exact cross-encoder scores for retrieved items~(line~\ref{alg_line:update_ce_scores}).
Since \adaCUR selects items uniformly at random in the first round,
\adaCUR with $\nRounds=1$ performs poorly as it simply returns a subset of items sampled uniformly at random.
As expected, Top-$k$-Recall for \adaCUR increases with the number of rounds and 
saturates at around 5-10 rounds while incurring negligible overhead in inference latency.
As shown in Figure~\ref{fig:rq_4_latency_breakdown_yugioh}, cross-encoder score computation is the main bottleneck in test-time inference, taking $\sim$7ms per 
score\footnote{For a \href{https://huggingface.co/nishantyadav/emb_crossenc_zeshel}{12-layer transformer model} on an Nvidia 2080ti}.
Increasing $\nRounds$ to large values such as 100 can incur up to 25\% overhead
with step (a) contributing significantly to this overhead.
Although the time taken by matrix multiplication in step (b) is \emph{linear} in
the number of items in the domain, we observe that it is a negligible 
fraction of overall latency on GPUs even for domain=\hotpotqa with 5 million 
items~(see~Figure~\ref{apndx_fig:rq_4_latency_breakdown_military_hotpotqa}) 
as GPUs
can enable significant speedup even for brute-force computation of this step.

%% file: figs/plots_rq_2.tex
\begin{figure*}[!ht]
     \includegraphics[width=.95\textwidth]{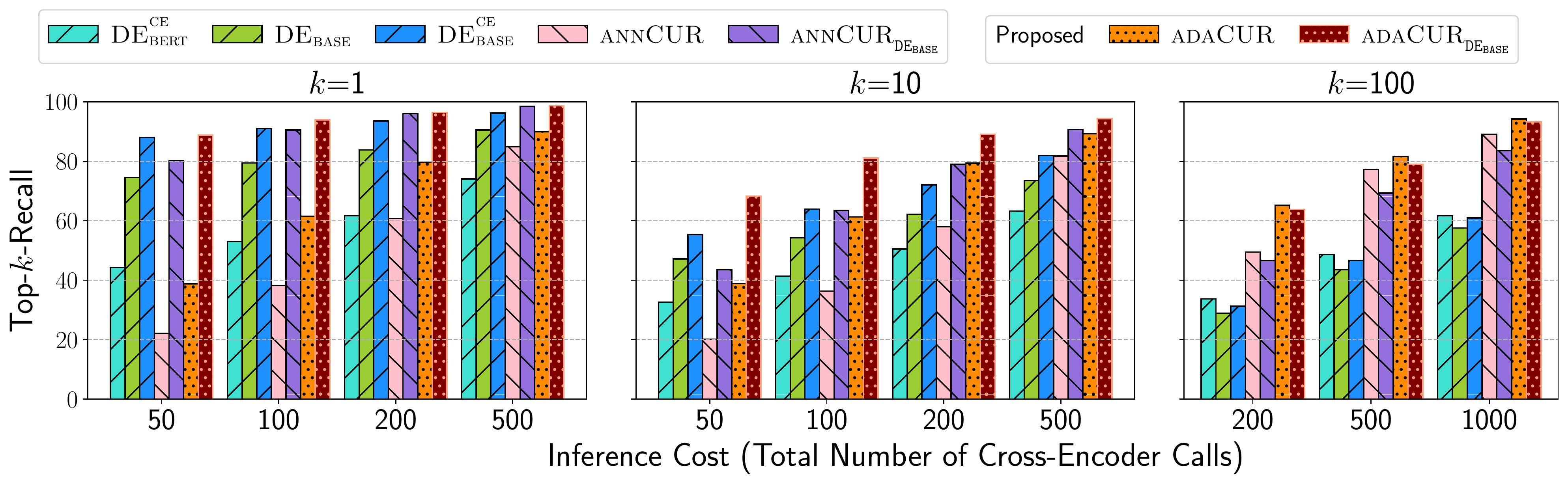}
    \caption{Top-$k$-Recall for \adaCUR and baselines for domain=\yugioh, $\queryTrainSize=500$. 
    \adaCUR  consistently outperforms the corresponding \annCUR variant and \adaCURwFixedDE 
    outperforms all DE-based retrieve-and-rerank approaches including 
    \finetuneDualEncoder, a DE model trained via distillation on the target domain using the cross-encoder.
    }
    \vspace{-0.2cm}
    \label{fig:rq_2_recall_at_same_cost_yugioh}
\end{figure*}

%% file: figs/plots_rq_3.tex
\begin{figure}[!ht]
    \centering
    
    \includegraphics[width=0.45\textwidth]{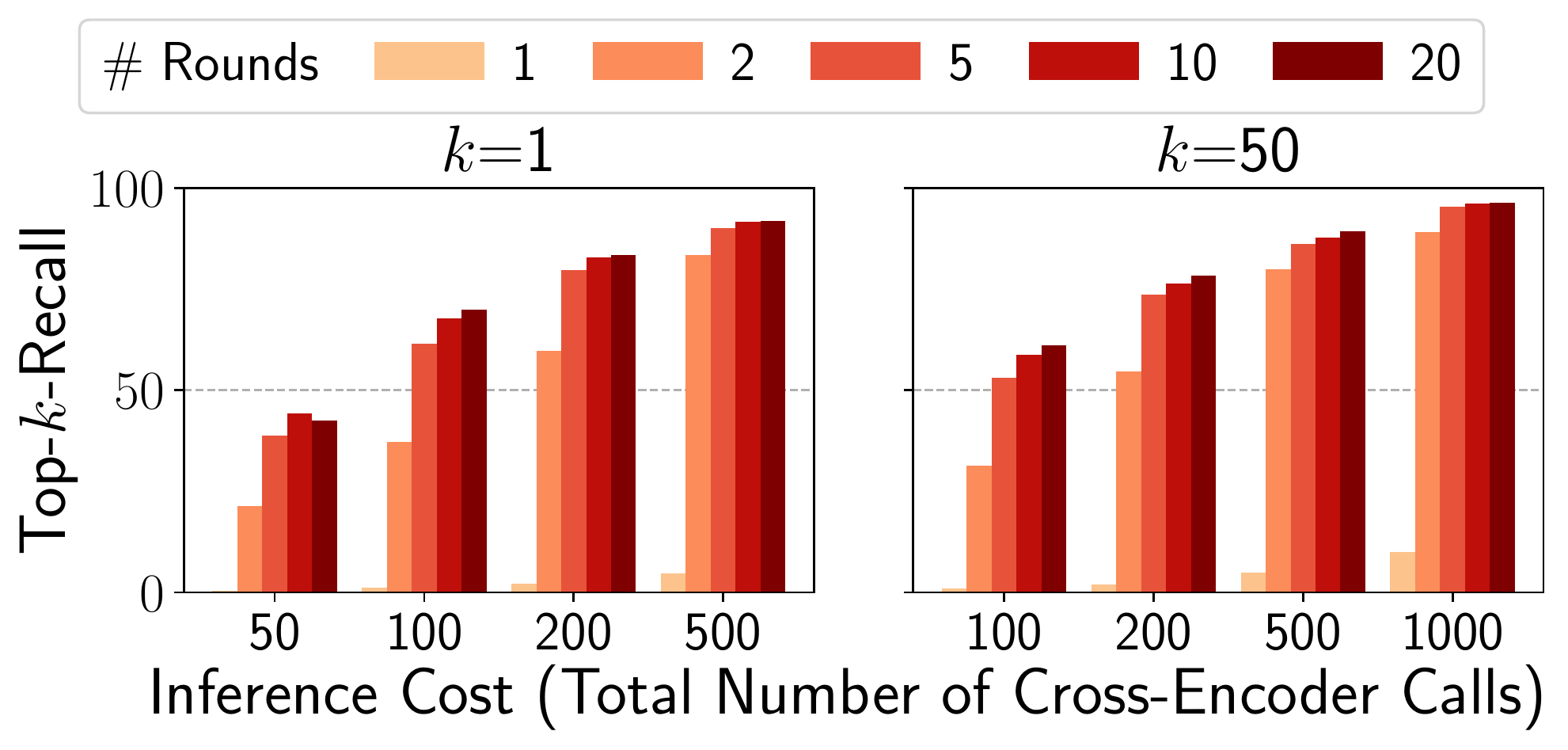}
    \caption{Top-$k$-Recall for \adaCUR for different number of rounds for domain=\yugioh, $\queryTrainSize=500$.
    }
    \label{fig:rq_3_recall_at_same_cost_yugioh_vs_n_steps}
\end{figure}

%% file: figs/plots_rq_4.tex
\begin{figure}[!ht]
     \centering 
     \includegraphics[width=0.49\textwidth]{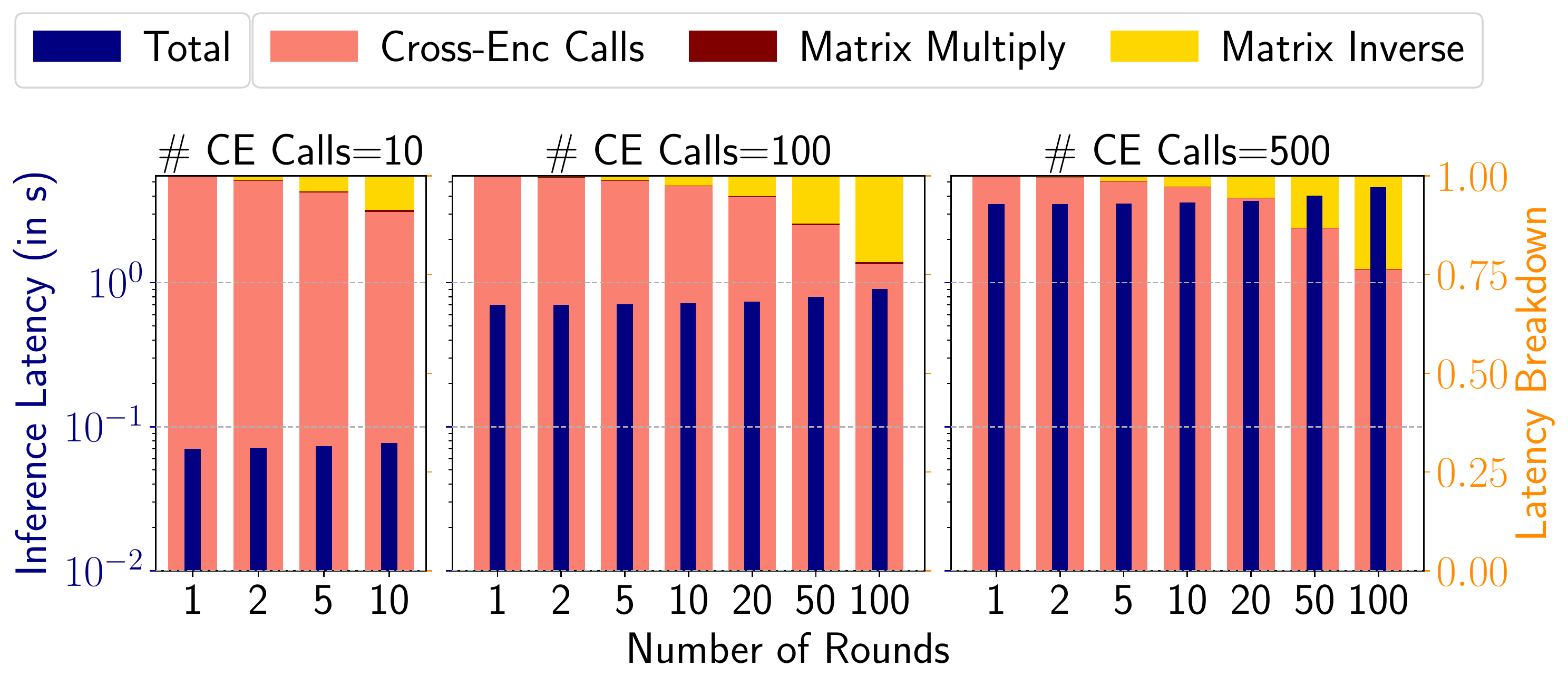}
    \vspace{-0.5cm}
    \caption{\adaCUR inference latency versus number of rounds for domain=\yugioh ,  $\queryTrainSize=500$.}
    \label{fig:rq_4_latency_breakdown_yugioh}
\end{figure}

%% file: tex/05_conclusion.tex
\section{Conclusion}

In this paper, we presented an adaptive search strategy that
incrementally builds a query embedding to approximate
cross-encoder scores and performs $k$-NN search using approximate
scores over several rounds.
Our approach is designed to reduce approximation error for the
top-scoring items and hence improves $k$-NN search recall
when retrieving items based on the approximate scores.
We perform an in-depth empirical analysis of the proposed approach  
in terms of both retrieval quality and efficiency.

%% file: tex/06_limitations_ethics.tex
\section*{Limitations}

Building the index for the \adaCUR is more expensive than the traditional dual-encoder index
due to the computation of dense cross-encoder scores matrix~(see \S\ref{subsec:adacur_indexing}).
We have successfully run our approach on up to 5 million items, but scaling to billions of items
is an interesting direction for future work.
Dual-encoder-based retrieve-and-rerank baseline approaches can benefit from training the
dual-encoder on multiple domains. It is not clear if data from multiple domains can be leveraged
to improve performance of the proposed approach on a given target domain; although in any case, cross-encoders tend to be more robust to domain shift than using only dual-encoders for retrieval.

\section*{Ethics Statement}
In this paper we tackle the task of finding $k$-nearest neighbor items for a given query
when query-items scores are computed using a black-box similarity function such as 
a cross-encoder model. 
The cross-encoder scoring function may have certain biases and error tendencies,
and it is unclear if our proposed method to approximate cross-encoder scores exacerbates
or mitigates such biases. 

%% file: tex/07_ack.tex
\section*{Acknowledgments}
We thank members of UMass IESL for helpful discussions
and feedback. 
This work was supported 
in part by the Center for Data Science and the Center
for Intelligent Information Retrieval, 
in part by the
National Science Foundation under Grant No. NSF1763618, 
in part by the Chan Zuckerberg Initiative under the project “Scientific Knowledge Base Construction”, 
in part by International Business Machines Corporation Cognitive Horizons Network agreement number W1668553, in part by
Amazon Digital Services, and 
in part using high-performance computing equipment obtained
under a grant from the Collaborative R\&D Fund
managed by the Massachusetts Technology Collaborative. 
Any opinions, findings, conclusions, and recommendations expressed in this material are
those of the authors and do not necessarily reflect those of
the sponsor(s).

%% file: tex/08_appendix.tex
\section{Appendix A}
\label{apndx_sec:appendix_A}

\input{tables_and_algos/data_stats.tex}

\section{Training and Implementation Details}
\label{apndx_sec:training_details}
All the code for reproducing experiments is available at \href{https://github.com/iesl/anncur}{https://github.com/iesl/anncur}.

\subsection{Training Cross-Encoder Models}
In our experiments, we use \eCrossenc, a cross-encoder model
variant proposed by~\citet{yadav2022efficient} that
jointly encodes a query-item pair and computes the final score
using dot-product of contextualized query and item embeddings
extracted after joint encoding.

\subsubsection{\zeshel Dataset}
\zeshel dataset is a zero-shot entity linking containing a set of 16 domains, 
each containing a disjoint set of items (entities).
Each domain contains a set of queries (mention) paired with their ground-truth items (entities).
For \zeshel, we use the cross-encoder model checkpoint\footnote{ \href{https://huggingface.co/nishantyadav/emb_crossenc_zeshel}{https://huggingface.co/nishantyadav/emb\_crossenc\_zeshel}} 
released by~\citet{yadav2022efficient}.
The cross-encoder model was trained by first
training a dual-encoder model on \zeshel training
data using hard negatives, and then training a cross-encoder
model for the task of zero-shot entity-linking on
all eight training domains using cross-entropy loss
with ground-truth entity and negative entities mined
using the dual-encoder. We refer readers to \citet{yadav2022efficient}
for further details on cross-encoder training.

We perform $k$-NN experiments on domains \yugioh, \starTrek, and \military
from \zeshel of which only \military was part of the training data
used to train the cross-encoder model and \yugioh and \starTrek
are part of the original \zeshel test domains and the cross-encoder
model was \emph{not} trained on these domains.

\subsubsection{\beir}

We follow the training setup used by~\citet{Hofsttter2020ImprovingEN}. We
first train three teacher cross-encoders  initialized
with \texttt{albert-large-v2}~\cite{lan2019albert}, 
\texttt{bert-large-whole-word-masking}, 
and \texttt{bert-base-uncased}~\cite{devlin-etal-2019-bert},
and compute soft labels on 40 million (query, positive item, negative item) triplets
in MS-MARCO dataset~\cite{bajaj2016ms}.
We then train our cross-encoder model
parameterized using a 6-layer \textsc{Mini-LM} model~\cite{wang2020minilm} via distillation
using average scores of the three teacher models as target signal
and minimizing mean-square-error between predicted and target scores.
We use training scripts
available as part of \texttt{sentence-transformer}\footnote{\href{https://github.com/UKPLab/sentence-transformers/blob/master/examples/training/ms_marco/train_cross-encoder_kd.py}{https://github.com/UKPLab/sentence-transformers}} repository
to train the cross-encoder model
and use a dot-product based scoring mechanism
for cross-encoders proposed by~\citet{yadav2022efficient}..

\input{figs/appendix_rq_4}

\subsection{Training Dual-Encoder Models}
\subsubsection{\zeshel dataset}
We report results for DE baselines as reported in~\citet{yadav2022efficient}.
The DE models were initialized using \texttt{bert-base-uncased} and
contain separate query and item encoders, thus containing a total of $2\times110M$ parameters.
We refer readers to~\citet{yadav2022efficient} for details related to 
the training of all DE model variants on \zeshel dataset.

\subsubsection{\beir benchmark}
For \beir domains, we use a dual-encoder model\footnote{\href{https://www.sbert.net/docs/pretrained-models/msmarco-v2.html}{\texttt{msmarco-distilroberta-base-v2}: www.sbert.net/docs/pretrained-models/msmarco-v2.html}}
released as part of \texttt{sentence-transformer} repository as \fixedDualEncoder. 
This dual-encoder model was initialized using 
\texttt{distillbert-base}~\cite{sanh2019distilbert} and trained 
on MS-MARCO dataset.
This \fixedDualEncoder is \emph{not} trained on target domains \scidocs and \hotpotqa
used for running $k$-NN experiments.

We finetune \fixedDualEncoder via distillation on the target domain to get \finetuneDualEncoder model.
Given a set of training queries $\queryTrainData$ from the target domain,
we retrieve top-100 or top-1000 items for each query, score the items with the cross-encoder
model, and train the dual-encoder by minimizing cross-entropy loss between
predicted query-item scores (using DE) and target query-item scores (obtained 
using cross-encoder).
Training a \finetuneDualEncoder with 1K queries and 100 or 1000 items per query takes
around 2 hrs and 10 hrs respectively on an Nvidia RTX8000 GPU with 48GB memory.
We train \finetuneDualEncoder for 10 epochs when using top-100 items per query and
for 4 epochs when using top-1000 items per query using AdamW~\cite{loshchilov2017decoupled} optimizer with learning
rate 1e-5.

\subsection{\annCUR Implementation details}
For \annCUR, we report results for the optimal split
of cross-encoder call budget ($\ceBudget$) between scoring $\nAnchorItems$ anchor items followed by
retrieving $\ceBudget-\nAnchorItems$ items for re-ranking.
We experiment with
$\nAnchorItems \in \{i\ceBudget/10 : 1 \leq i \leq 9\}$.
If the retrieved items contain a subset of anchor
items for which exact cross-encoder score has already been computed, we retrieve more than $\ceBudget-\nAnchorItems$ items using approximate scores and compute exact cross-encoder scores for them until we have exhausted the entire
cross-encoder call budget for the re-ranking step.

\subsection{\adaCUR Implementation details}
For all our $k$-NN search experiments, we used Nvidia 2080ti GPUs with 12GB memory
for domains \yugioh (10K items), \starTrek (34K items), \military (100K items), and \scidocs (25K items), and we used Nvidia RTX8000 GPUs
with 48GB memory for \hotpotqa (5 million items).

For \hotpotqa, we restrict our $k$-NN search to
top-10K items wrt $\fixedDualEncoder$ for $\adaCURwFixedDE$.
For \zeshel domains and \scidocs, we do not use any such heuristic and search over all 
the items in the corresponding domain.

\subsection{Time Complexity of \adaCUR}
\label{apndx_subsec:time_complexity}

The offline indexing step for \adaCUR takes 
$\mathcal{O}(\nAnchorQueries \nItems \cost_{\model{\theta}})$ time as 
it involves computing exact cross-encoder scores
for all $\nItems$ items in the target domain against $\nAnchorQueries$ anchor queries, and computing each cross-encoder score takes $\cost_{\model{\theta}}$ units of time.

At test time, we are given a budget $\ceBudget$ on the number of cross-encoder calls.
Each one of the $\nRounds$ rounds during inference with \adaCUR involves approximating 
all item scores for the test query ($\testQuery$) followed by sampling 
the next batch of $\nAnchorItemsPerRound=\ceBudget/\nRounds$ items using the updated approximate scores.
In the $\roundIter^{th}$ round, the score approximation step involves computing
the pseudo-inverse of a $\nAnchorQueries\times \roundIter\nAnchorItemsPerRound~$-dimensional 
matrix~(line~\ref{alg_line:approx_score_cur_inv} in Algo.~\ref{alg:incremental_cur}),
which takes $\mathcal{O}(\cost_\texttt{inv}^{\nAnchorQueries, \roundIter\nAnchorItemsPerRound})$ time,
followed by a matrix multiplication step to compute updated approximate
scores~(line~\ref{alg_line:approx_score_cur_mat_mul} in Algo.~\ref{alg:incremental_cur})
which takes $\mathcal{O}(\cost_\texttt{mul}^{\nAnchorQueries, \roundIter\nAnchorItemsPerRound, \nItems})$ time.
The time taken to update the approximate scores in each round is 
$\mathcal{O}(\cost_\texttt{inv}^{\nAnchorQueries,\roundIter\nAnchorItemsPerRound}+\cost_\texttt{mul}^{\nAnchorQueries,\roundIter\nAnchorItemsPerRound, \nItems})$,
and the time taken to compute cross-encoder scores for the next batch of $\nAnchorItemsPerRound$ items
is $\mathcal{O}(\nAnchorItemsPerRound \cost_{\model{\theta}})$.
Thus, the total inference latency for retrieving items over $\nRounds$ rounds 
under a given budget of $\ceBudget$ cross-encoder calls is
\begin{align*}
    &\mathcal{O}\Big( \sum_{j=1}^{\nRounds} \big( \nAnchorItemsPerRound \cost_{\model{\theta}} + \cost_\texttt{inv}^{\nAnchorQueries, \roundIter\nAnchorItemsPerRound} + \cost_\texttt{mul}^{\nAnchorQueries, \roundIter\nAnchorItemsPerRound, \nItems} \big) \Big)   \\
    =& \mathcal{O}\Big(  \ceBudget\cost_{\model{\theta}} + \underbrace{\sum_{j=1}^{\nRounds} \big( \cost_\texttt{inv}^{\nAnchorQueries, \roundIter\nAnchorItemsPerRound} + \cost_\texttt{mul}^{\nAnchorQueries, \roundIter\nAnchorItemsPerRound, \nItems} \big) }_{\text{Overhead of \adaCUR}} \Big) 
\end{align*}

Figure~\ref{apndx_fig:rq_4_latency_breakdown_military_hotpotqa} shows 
the breakdown of \adaCUR's inference latency in terms of time spent on 
computing cross-encoder scores, and the overhead of computing matrix inverse in line~\ref{alg_line:approx_score_cur_inv} and updating approximate scores by multiplying matrices in line~\ref{alg_line:approx_score_cur_mat_mul} of Algorithm~\ref{alg:incremental_cur}.
Empirically, we observe that the primary bottleneck at 
inference time is the time taken to compute cross-encoder scores for query-item pairs at test time, and the overhead for \adaCUR
accumulates linearly as the number of rounds increases. The overhead is mostly dominated by 
computing pseudo-inverse~(see line~\ref{alg_line:approx_score_cur_inv} in Algorithm~\ref{alg:incremental_cur})
and this step is independent of the target domain size. 
The matrix multiplication step~(line~\ref{alg_line:approx_score_cur_mat_mul} in Algorithm~\ref{alg:incremental_cur}) has a linear dependence on
the number of items in the target domain but it is a negligible fraction of the overall running time as it can be significantly
sped up using GPUs. 

For \zeshel domains, we use a cross-encoder parameterized using
\texttt{bert-base}~\cite{devlin-etal-2019-bert}, and observe that
each cross-encoder call takes amortized time of $\sim$7ms on an Nvidia 2080ti GPU when the scores are computed in batches of size 50.
Computing each cross-encoder score sequentially i.e. with batch-size = 1 takes $\sim$13ms per score. We did not observe any further reduction in
amortized time to compute each score when increasing the batch size beyond 50.

The amortized time per cross-encoder call is approximately 6ms and 2ms for \scidocs and \hotpotqa respectively when using batch size=50 and \textsc{Mini-LM}-based~\cite{wang2020minilm} cross-encoder. 
The difference in time per cross-encoder score for \scidocs and \hotpotqa is due
to the difference in average query-item pair sequence length.

\section{Additional Results and Analysis}
\label{apndx_sec:additional_results}

\subsection{\topkSample vs \softmaxSample for \adaCUR}
Figure~\ref{apndx_fig:rq_9_topk_vs_softmax_adacur} shows Top-$k$-Recall for \adaCUR on domain=\yugioh,
$\lvert \queryTrainData \rvert = 500$, when using 
\topkSample and \softmaxSample strategies for sampling items based
on approximate scores~(see~\S\ref{subsec:adacur_inference} for details).
\topkSample sampling strategy which greedily picks top-$k$ items 
based on approximate scores results in superior recall as
compared to sampling items using softmax over approximate scores.

\input{figs/appendix_rq_9_topk_vs_softmax}

\subsection{Anchor Item Sampling with Oracle}
\label{apndx_subsec:oracle_exp}
\adaCUR performs retrieval over multiple rounds 
using  approximate cross-encoder scores and uses the
items retrieved based on the approximate scores as
anchor items to improve the approximation and hence retrieval in subsequent rounds.
In this section, we run experiments where the 
anchor item sampling method has \emph{oracle} access to exact cross-encoder scores 
of all items for the given test query 
to better understand the effect of anchor items on the approximation
of cross-encoder scores and hence subsequent retrieval based on the approximate scores.
We experiment with the following strategies for sampling $\nAnchorItems$ anchor items for
a given test query :
\begin{itemize}[topsep=1pt,itemsep=0ex,partopsep=1ex,parsep=1ex]
    \item $\epsTopkGTSample{\Kmask}{\epsilon}$ : Mask out top-$\Kmask$ items wrt exact cross-encoder scores and select $\nAnchorItems$ anchor items by  greedily picking $(1-\epsilon)\nAnchorItems$ items starting from rank $\Kmask+1$, and sample remaining $\epsilon \nAnchorItems$ anchor items uniformly at random.
    \item $\epsSoftmaxGTSample{\Kmask}{\epsilon}$: Mask out top-$\Kmask$ items wrt exact cross-encoder scores and select $\nAnchorItems$ anchor items
    by sampling $(1-\epsilon)\nAnchorItems$ anchor items using softmax over exact cross-encoder scores, and sample remaining $\epsilon \nAnchorItems$ anchor items uniformly at random. 
\end{itemize}

For a given test-time cross-encoder call budget $\ceBudget$, we 
select $\nAnchorItems$ anchor items, compute approximate cross-encoder
scores using the chosen anchor items, and then retrieve $\ceBudget - \nAnchorItems$
items based on the approximate scores. 
We experiment with 
$\nAnchorItems \in \{i\ceBudget/10 : 1 \leq i \leq 9\}$ and report results for the best budget split.

\input{figs/plots_rq_7.tex}

\input{figs/plots_rq_6.tex}

\input{figs/plots_rq_6_examples.tex}

\paragraph{Effect of adding $k$-NN items to anchor items } 
Figure~\ref{fig:rq_6_exact_sampling_w_wo_topk_yugioh} shows Top-$k$-Recall of 
anchor item sampling strategies $\epsTopkGTSample{\Kmask}{0}$ and 
$\epsSoftmaxGTSample{\Kmask}{0}$ for $\Kmask \in \{0,k\}$,  domain=\yugioh.
Sampling strategies $\epsSoftmaxGTSample{k}{0}$ 
and $\epsTopkGTSample{k}{0}$, which mask out top-$k$ items, 
perform significantly worse than $\epsSoftmaxGTSample{0}{0}$  and 
$\epsTopkGTSample{0}{0}$ respectively when searching 
for $k=1,10$ nearest neighbors.
This indicates that the significant improvement in Top-$1$-Recall and Top-$10$-Recall
for $\epsTopkGTSample{0}{0}$ and $\epsSoftmaxGTSample{0}{0}$ sampling strategies
can be attributed to the presence of 
top-$k$ items in the anchor item set.
This is because CUR matrix factorization which is used to compute the approximate
scores incurs negligible approximation error on anchor items,
and hence on top-$k$ items when these items are part of the anchor set
as shown in figures~\ref{fig:rq_7_error_yugioh} and \ref{fig:rq_6_example_exact_sampling_method_yugioh}.
For $\epsTopkGTSample{k}{0}$ and $\epsSoftmaxGTSample{k}{0}$ sampling strategies, 
since the top-$k$ items are \emph{not} part of the anchor set,
CUR incurs a much higher approximation error for the 
top-$k$ items~(see examples in Figures~\ref{fig:rq_6_example_exact_sampling_greedy_wo_topk_yugioh} and 
\ref{fig:rq_6_example_exact_sampling_smax_wo_topk_yugioh} ), thus resulting in poor Top-$k$-Recall
as shown in Figure~\ref{fig:rq_6_exact_sampling_w_wo_topk_yugioh}.

\paragraph{Effect of \emph{diversity} in anchor items}
Figure~\ref{fig:rq_6_exact_sampling_w_wo_topk_yugioh} shows
that sampling items based on softmax of exact cross-encoder scores~($\epsSoftmaxGTSample{\Kmask}{0}$) 
performs better than greedily picking top-scoring 
items~($\epsTopkGTSample{\Kmask}{0}$), for both $\Kmask=0,k$.
The reason behind $\epsSoftmaxGTSample{\Kmask}{0}$ performing better than $\epsTopkGTSample{\Kmask}{0}$ is that sampling based on
softmax of exact scores yields an anchor set with a more
\emph{diverse} score distribution whereas 
greedily selecting top-scoring 
items using exact scores results in an anchor set with items 
having similar cross-encoder scores. 
However, as shown in 
Figures~\ref{fig:rq_6_example_exact_sampling_greedy_yugioh} and~\ref{fig:rq_6_example_exact_sampling_smax_yugioh},
both of these sampling strategies can result in 
overestimating scores for all items, even the irrelevant ones (i.e. items beyond
top-$k$ items)
due to insufficient representation of the irrelevant items in
the anchor set.
Thus retrieving based on approximated scores may struggle
to retrieve relevant $k$-NN items, especially for larger values of $k$ such as $k=100$
when the anchor items are chosen using oracle strategies such as $\epsTopkGTSample{\Kmask}{0}$.

Figures~\ref{fig:rq_6_example_exact_sampling_greedy_rand_eps_75_yugioh} and ~\ref{fig:rq_6_example_exact_sampling_smax_rand_eps_75_yugioh}, where $\epsilon=75\%$ of 50 
items are sampled uniformly at random, show that 
overestimating scores of irrelevant items can be avoided
by sampling a fraction of anchor items uniformly at random 
to increase the diversity of the anchor item set.
As shown in Figures~\ref{fig:rq_6_exact_sampling_greedy_w_rand_yugioh}  
and~\ref{fig:rq_6_exact_sampling_smax_w_rand_yugioh}, 
Top-$k$-Recall for both $\epsSoftmaxGTSample{0}{\epsilon}$ and 
$\epsTopkGTSample{0}{\epsilon}$ generally improves 
with an increase in $\epsilon$, the fraction of random items in the anchor set, 
due to increased diversity in the anchor item set. 
Since $\epsSoftmaxGTSample{0}{\epsilon}$ already 
samples a diverse set of anchor items, increasing $\epsilon$ 
yields only marginal improvement while for 
$\epsTopkGTSample{0}{\epsilon}$, increasing $\epsilon$ yields significant
improvements due to increased diversity of the anchor set.
A small drop in performance is observed
for larger values of $\epsilon$ as increasing $\epsilon$ beyond a threshold
results in some of the top-$k$ items to be excluded from the anchor item set.
This results in a poorer approximation of scores for the missing top-$k$ items
and hence poor retrieval recall as the retrieval is done using the approximate scores.

Finally, the optimal strategy for choosing the set of anchor items 
is the one that strikes a balance between selecting
anchor items with diverse cross-encoder scores and greedily picking top-$k$ items.
Our proposed strategy \adaCUR improves over \annCUR as 
greedily picking top-scoring items according to approximate scores 
to expand set of anchor items increases the likelihood of picking 
ground-truth $k$-NN items to be part of the anchor set, 
with this likelihood improving after each round with improvement
in the score approximation,
and \adaCUR achieves diversity in the anchor items as a result of sampling
items uniformly at random in the first round and due to error in the approximate scores,
as shown in Figure~\ref{apndx_fig:example_icur_sampling_per_round}.

\subsection{Comparison with Multi-Vector Models} 
\label{apndx_subsec:multi_vec_results}
Multi-vector models~\cite{khattab2020colbert, ma-etal-2021-muver} produce multiple embeddings for each 
query and item. For a given query $\query$ and item $\dataItem$, 
the query-item score is computed using simple functions such
as average similarity or sum-of-maximum similarities
between the set of embeddings for query $\query$ and item $\dataItem$.

\input{figs/appendix_rq_10_multi_vec}

Figure~\ref{apndx_fig:rq_10_multi_vec_zeshel} shows Top-$k$-Recall for \fixedDualEncoder, \finetuneDualEncoder, \adaCURwFixedDE,  and \textsc{Muver}~\cite{ma-etal-2021-muver},
a recent multi-vector model trained on \zeshel dataset.
For \muver, we use the pre-trained checkpoint released by~\citet{ma-etal-2021-muver}
with the \emph{view-merging} inference strategy as described in~\citet{ma-etal-2021-muver}.
While \textsc{Muver} can be more accurate than \fixedDualEncoder, \finetuneDualEncoder obtained by finetuning \fixedDualEncoder model on the target domain outperforms \muver and our proposed method
\adaCURwFixedDE yields the best Top-$k$-Recall versus inference cost trade-offs for all values of $k$.

We would also like to note that while multi-vector models such as \muver can be more accurate than single-embedding models such as \fixedDualEncoder, such multi-vector models incur significant memory overhead for storing query/item embeddings.
For instance, using 15 embeddings per item with 768-dimensional embeddings would take around 250GB space for 5 million items for \hotpotqa.

\subsection{Results for \tfidf baseline}
\label{apndx_subsec:tfidf_results}
\noindent \textbf{\tfidf}: All queries and items are embedded using a \tfidf vectorizer trained on item descriptions and top-$k$ items are retrieved using the dot-product of sparse query and item embeddings.

For domains in \zeshel, we report results for \tfidf baseline, for
\annCUR when anchor items are chosen using \tfidf baseline (\annCURwTFIDF),
and for \adaCUR when the first batch of anchor items
is chosen using \tfidf baseline (\adaCURwTFIDF).
Figures~\ref{apndx_fig:rq_2_recall_at_same_cost_yugioh},~\ref{apndx_fig:rq_2_recall_at_same_cost_starTrek},
and~\ref{apndx_fig:rq_2_recall_at_same_cost_military} show Top-$k$-Recall for domains 
\yugioh, \starTrek, and \military respectively for 
$\queryTrainSize \in \{100, 500, 2000\}$.
For each baseline retrieval method, \adaCUR always performs better
than \annCUR which in turn generally performs better than merely re-ranking items 
retrieved using the corresponding baseline retrieval method. 
In most cases, Top-$k$-Recall for 
\adaCURwFixedDE > \annCURwFixedDE > \fixedDualEncoder, and 
\adaCURwTFIDF > \annCURwTFIDF > \tfidf.

\input{figs/appendix_per_round_scatter.tex}

\input{figs/appendix_rq_2_beir}

\input{figs/appendix_rq_2.tex}

%% file: tables_and_algos/data_stats.tex
\begin{table*}[!ht]
    \small
    \centering
    \begin{tabular}{l l|c c c c }
        \toprule
        Dataset & Domain                  &  $\nItems$ & $(\lvert \queryTrainData \rvert, \lvert \queryTestData \rvert)$ Splits & Train Query~($\queryTrainData$) Type \\
        \midrule
        \zeshel  & \yugioh                &   10,031     & (100/3274), (500/2874), (2000/1374) & Real Queries\\
        \zeshel  & \starTrek              &   34,430     & (100/4127), (500/3727), (2000/2227) & Real Queries\\
        \zeshel  & \military               &   104,520    & (100/2300), (500/1900), (2000/0400) & Real Queries\\
        \midrule
        \beir  & \scidocs               &   25,657       & (1000/1000) & Pseudo-Queries \\
        \beir  & \hotpotqa              &    5,233,329  & (1000/1000)  & Pseudo-Queries \\
        \bottomrule
    \end{tabular}
    \caption{Statistics on the number of items ($\itemSpace$) and the number of queries in train and test splits for each domain. The train-query~($\queryTrainData$) split refers to queries used for distilling dual-encoder models or for indexing items using \adaCUR and \annCUR. For \zeshel domains, we create train-test splits by
    splitting the queries in each domain uniformly at random and test with three different splits by putting
    100, 500, or 2000 queries in train split. For \beir domains, we use pseudo-queries released as part of the
    benchmark as train queries~($\queryTrainData$) and run $k$-NN evaluation on test queries from the \emph{official} test split~(as per \beir benchmark) of these domains. For \hotpotqa, we use the first 1K queries out of a total of 7K test queries and we use all 1K test queries for \scidocs.
    }
    \label{tab:dataset_stats}
\end{table*}

%% file: figs/appendix_rq_4.tex
\begin{figure*}
    \centering
    \begin{subfigure}[b]{0.49\textwidth}
         \includegraphics[width=\textwidth]{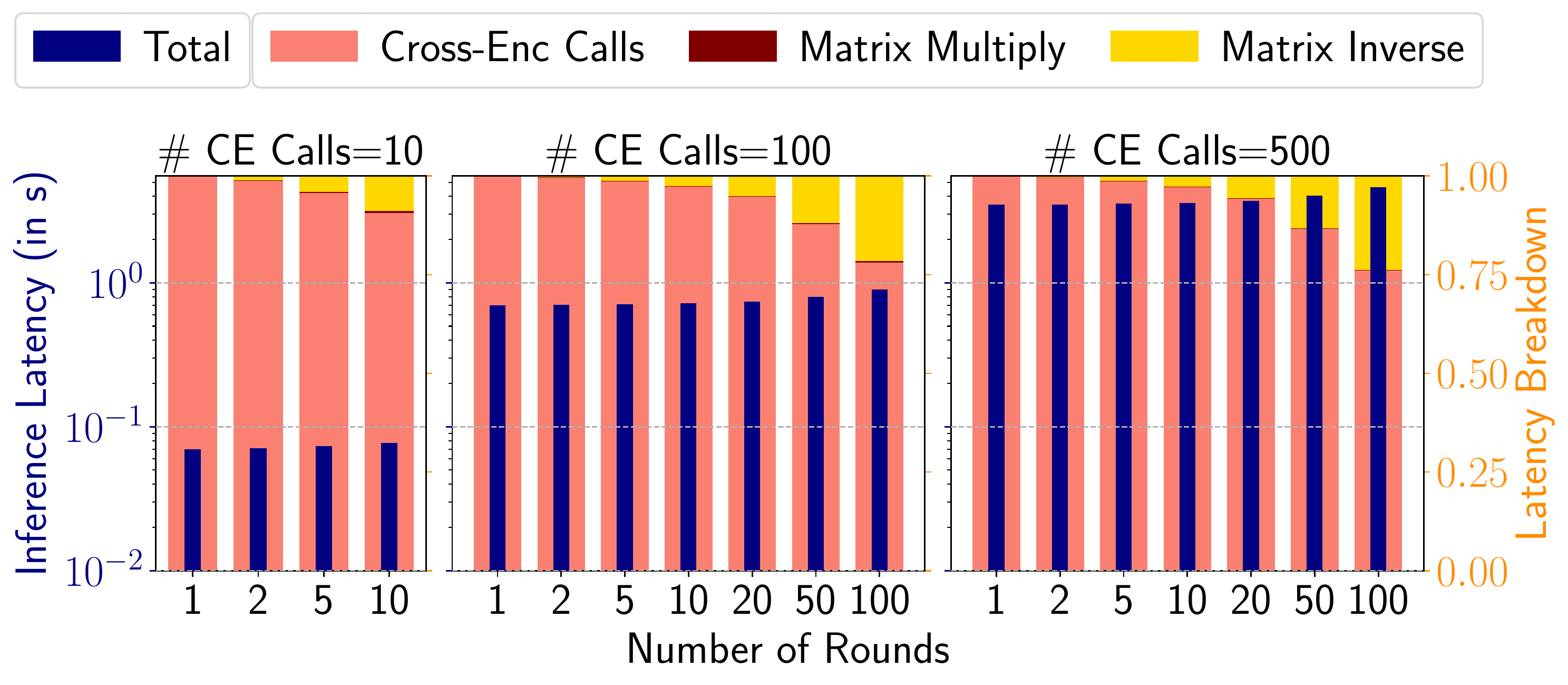}
        
        \caption{Domain=\military (100K items),  $\queryTrainSize=500$. Each CE call takes $\sim$7ms on an Nvidia 2080ti GPU
        for a cross-encoder parameterized using \href{https://huggingface.co/nishantyadav/emb_crossenc_zeshel}{12-layered transformer model}.
        }
        \label{apndx_fig:rq_4_latency_breakdown_military}
    \end{subfigure}
    \hfill
    \begin{subfigure}[b]{0.49\textwidth}
         \includegraphics[width=\textwidth]{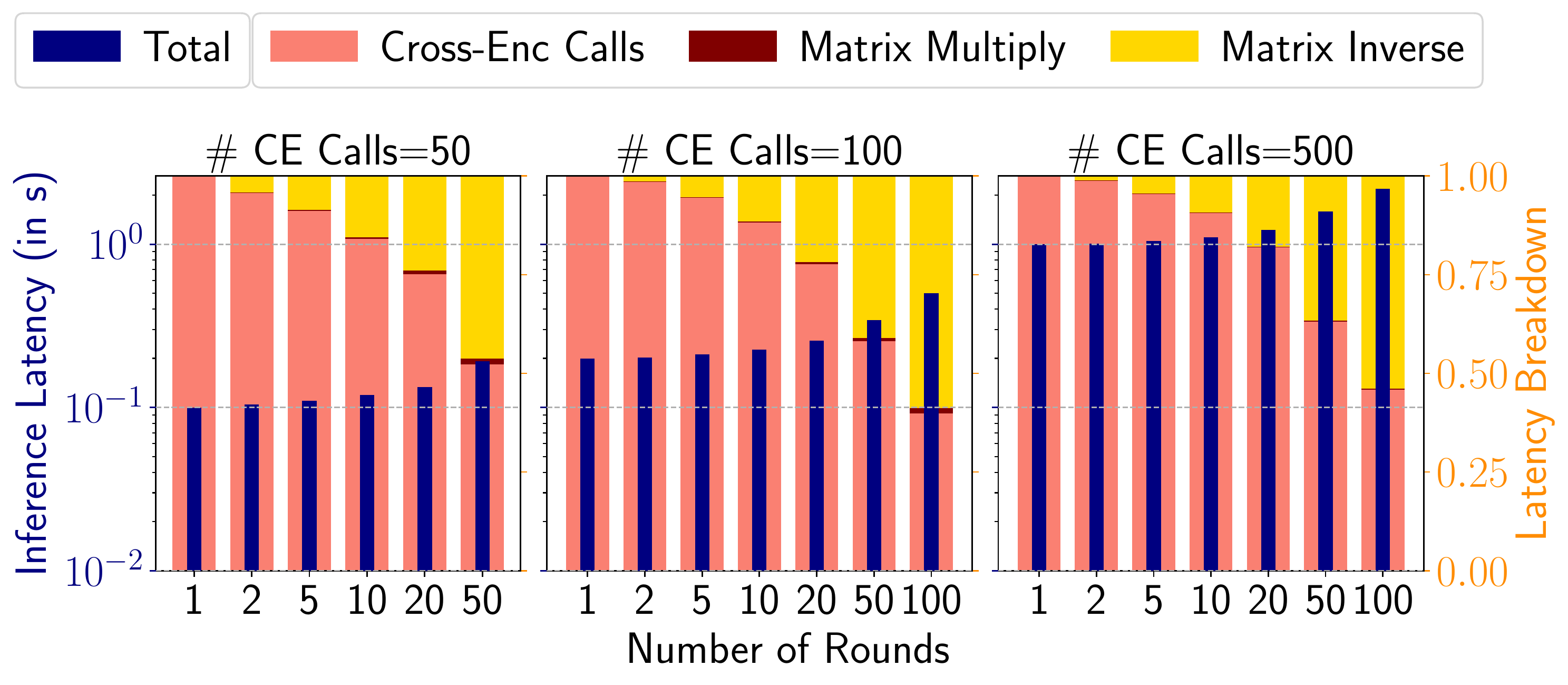}
        \caption{Domain=\hotpotqa (5 Million items),  $\queryTrainSize=1000$. Each CE call takes $\sim$2ms on an Nvidia RTX8000 GPU
        for a cross-encoder parameterized using \href{https://huggingface.co/cross-encoder/ms-marco-MiniLM-L-6-v2}{6-layered transformer model}.
        }
        \label{apndx_fig:rq_4_latency_breakdown_hotpotqa}
    \end{subfigure}
    
    \caption{\adaCUR inference latency versus number of rounds for two different domains. The primary bottleneck at 
    inference time is the time taken to compute cross-encoder~(CE) scores for query-item pairs at test time, and the overhead for \adaCUR
    accumulates linearly as the number of rounds increases. 
    See \S\ref{apndx_subsec:time_complexity} for detailed discussion.
    }
    \label{apndx_fig:rq_4_latency_breakdown_military_hotpotqa}
    
\end{figure*}

%% file: figs/appendix_rq_9_topk_vs_softmax.tex
\begin{figure}[!h]
    
    \includegraphics[width=0.48\textwidth]{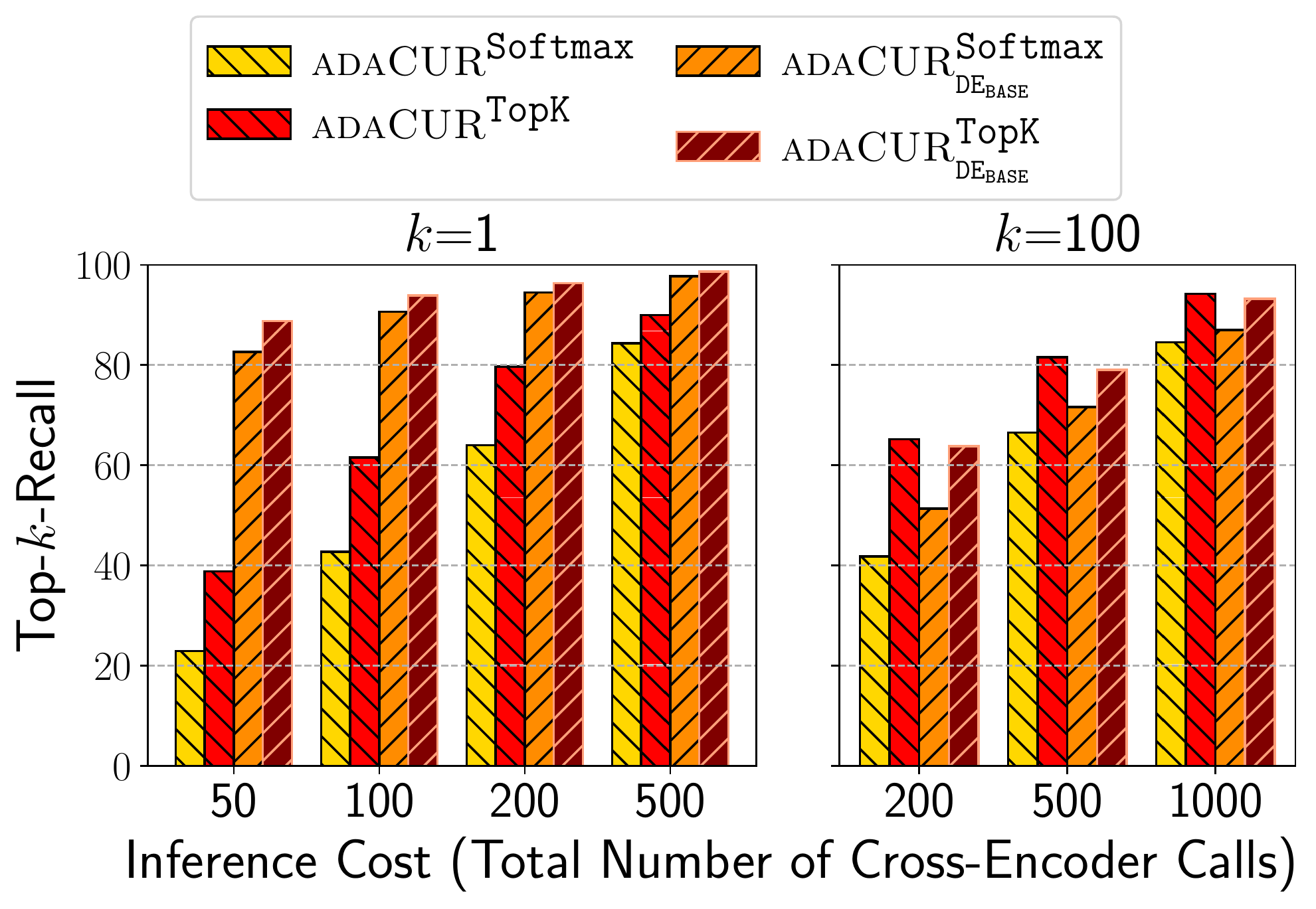}
        
    \caption{Top-$k$-Recall for \adaCUR on \yugioh, $\queryTrainSize=500$ for different strategies for sampling items based
    on approximate scores described in~\S\ref{subsec:adacur_inference}.
    }
    \label{apndx_fig:rq_9_topk_vs_softmax_adacur}
    \vspace{-0.5cm}
\end{figure}

%% file: figs/plots_rq_7.tex
\begin{figure}[!ht]
    \centering
     \includegraphics[width=0.46\textwidth]{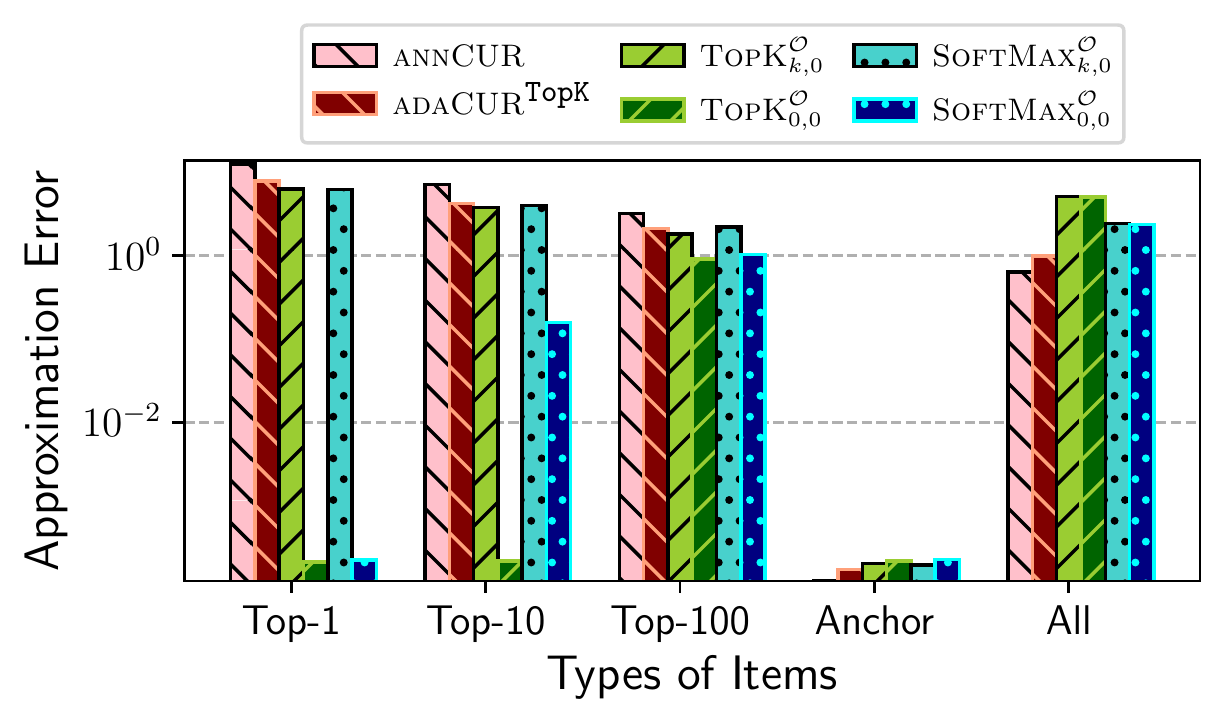}
     \vspace{-0.5cm}
    \caption{
    Average approximation error for CUR matrix factorization on test-queries for domain=\yugioh and
    $\queryTrainSize=500$ when choosing $\nAnchorItems=50$ anchor items uniformly at random~(\annCUR), using oracle strategies from~\S\ref{apndx_subsec:oracle_exp} and for \adaCUR when sampling anchor items over five rounds. Approximation error is computed as
    the average of absolute difference between approximate and exact item scores.
    }
    \label{fig:rq_7_error_yugioh}
    
\end{figure}

%% file: figs/plots_rq_6.tex
\begin{figure}[!t]
    \centering
    \begin{subfigure}[b]{0.48\textwidth}
         \includegraphics[width=\textwidth]{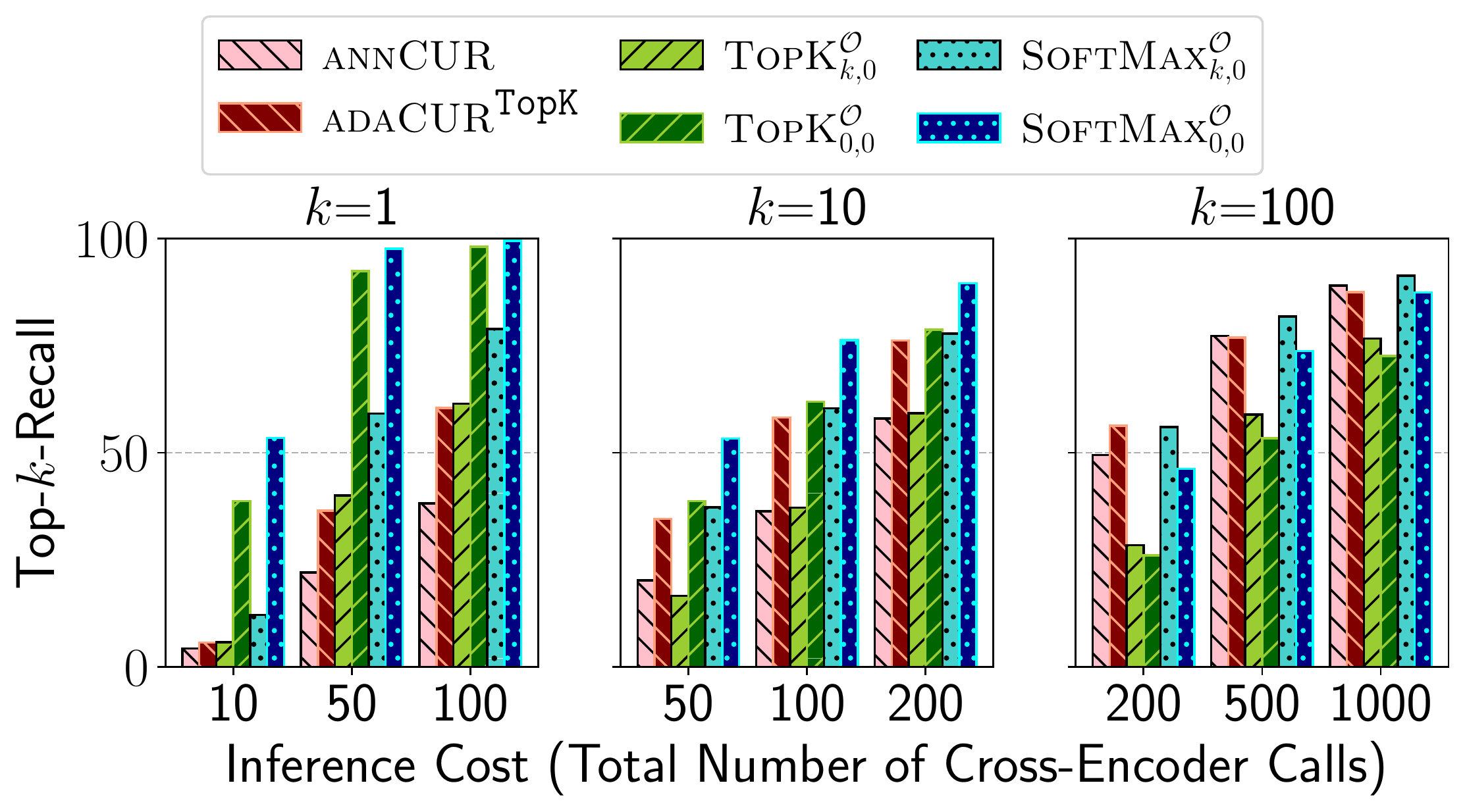}
        \caption{Sampling with and without ground-truth top-$k$ items}  
        \label{fig:rq_6_exact_sampling_w_wo_topk_yugioh}
    \end{subfigure}
     \hfill
    \begin{subfigure}[b]{0.48\textwidth}
         \includegraphics[width=\textwidth]{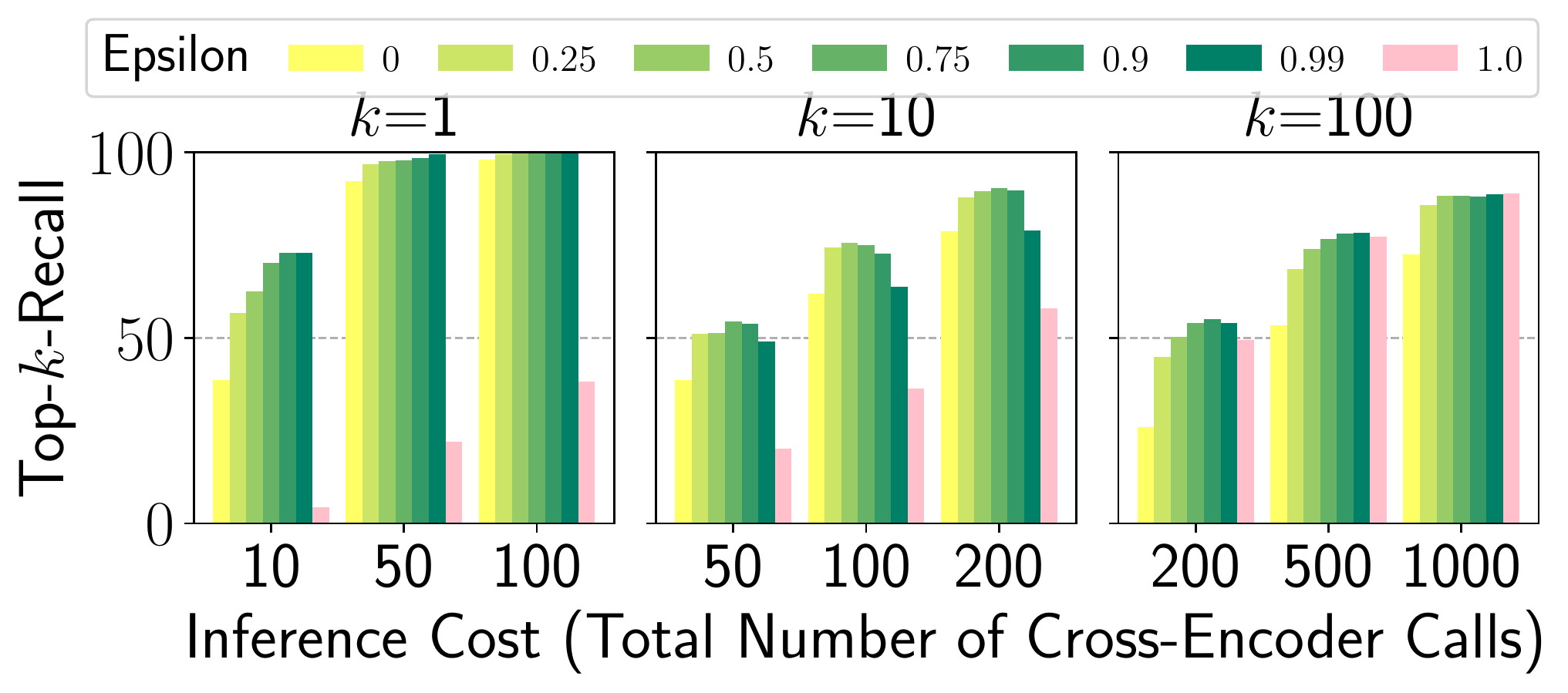}
        \caption{Sampling using \topkSample strategy with exact CE scores while varying $\epsilon$, the fraction of items sampled uniformly at random}  
        \label{fig:rq_6_exact_sampling_greedy_w_rand_yugioh}
    \end{subfigure}
     \hfill
          \begin{subfigure}[b]{0.48\textwidth}
         \includegraphics[width=\textwidth]{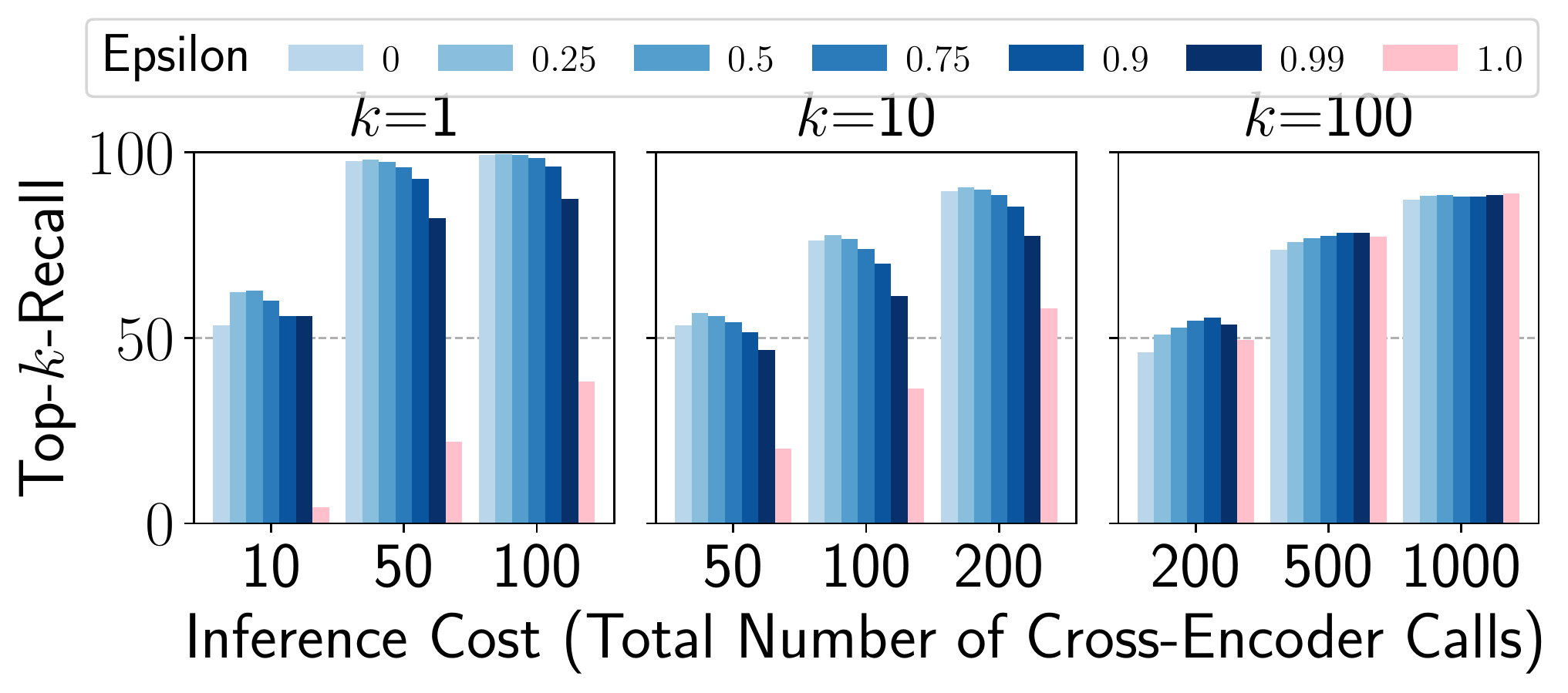}
        \caption{Sampling using \softmaxSample of exact CE scores while varying $\epsilon$, the fraction of items sampled uniformly at random}  
        \label{fig:rq_6_exact_sampling_smax_w_rand_yugioh}
    \end{subfigure}
    \caption{
    Top-$k$-Recall for \adaCUR, \annCUR, and oracle sampling strategies~(\S\ref{apndx_subsec:oracle_exp}) that have oracle access to exact cross-encoder scores for all items
    for domain=\yugioh, $\queryTrainSize=500$.
    }
    \label{fig:rq_6_icur_sampling_method_yugioh}
    \vspace{-0.6cm}
\end{figure}

%% file: figs/plots_rq_6_examples.tex
\begin{figure}[!ht]
    \centering
    \begin{subfigure}[b]{0.4\textwidth}
         \includegraphics[width=\textwidth]{figs/scatter_plots/legend.pdf}
        \phantomcaption{} 
        \vspace{-0.5cm}
    \end{subfigure}
    \addtocounter{subfigure}{-1}
     \begin{subfigure}[b]{0.23\textwidth}
         \includegraphics[width=\textwidth]{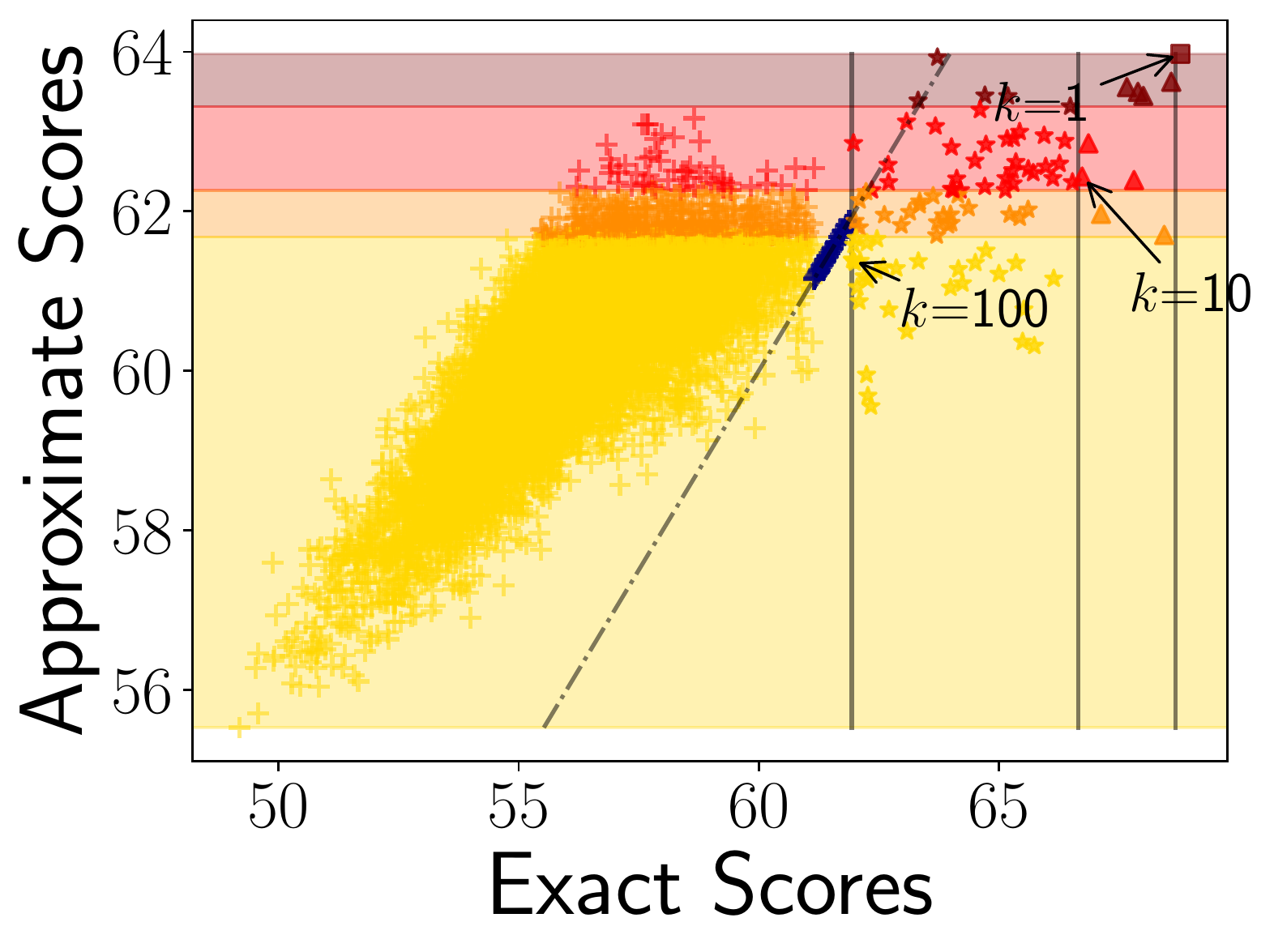}
        \caption{$\epsTopkGTSample{100}{\epsilon}$, $\epsilon=0.0$}  
        \label{fig:rq_6_example_exact_sampling_greedy_wo_topk_yugioh}
    \end{subfigure}
    \hfill
    \begin{subfigure}[b]{0.23\textwidth}
         \includegraphics[width=\textwidth]{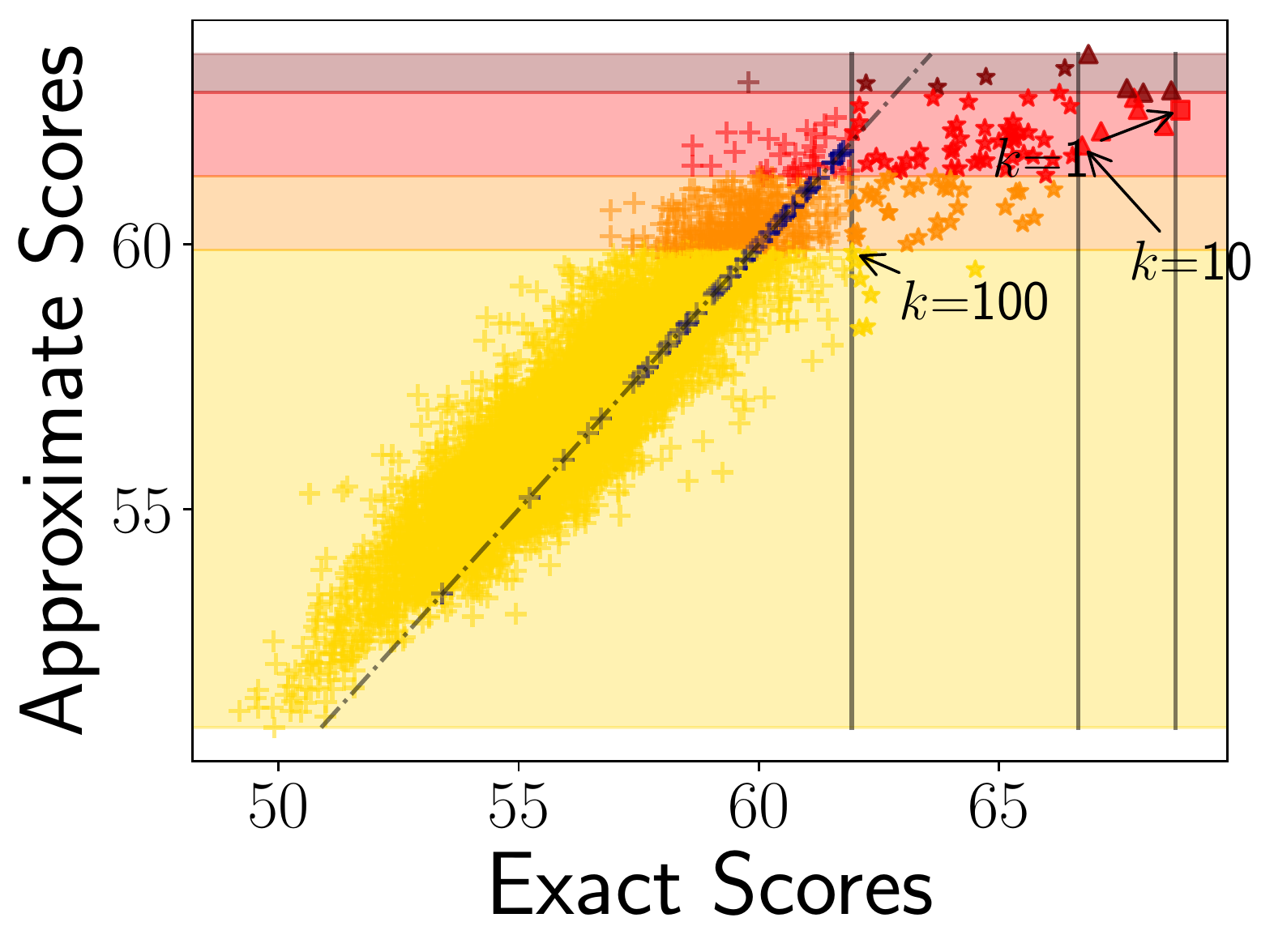}
        \caption{$\epsSoftmaxGTSample{100}{\epsilon}$, $\epsilon=0.0$}
         \label{fig:rq_6_example_exact_sampling_smax_wo_topk_yugioh}
    \end{subfigure}
    \hfill
    \begin{subfigure}[b]{0.23\textwidth}
         \includegraphics[width=\textwidth]{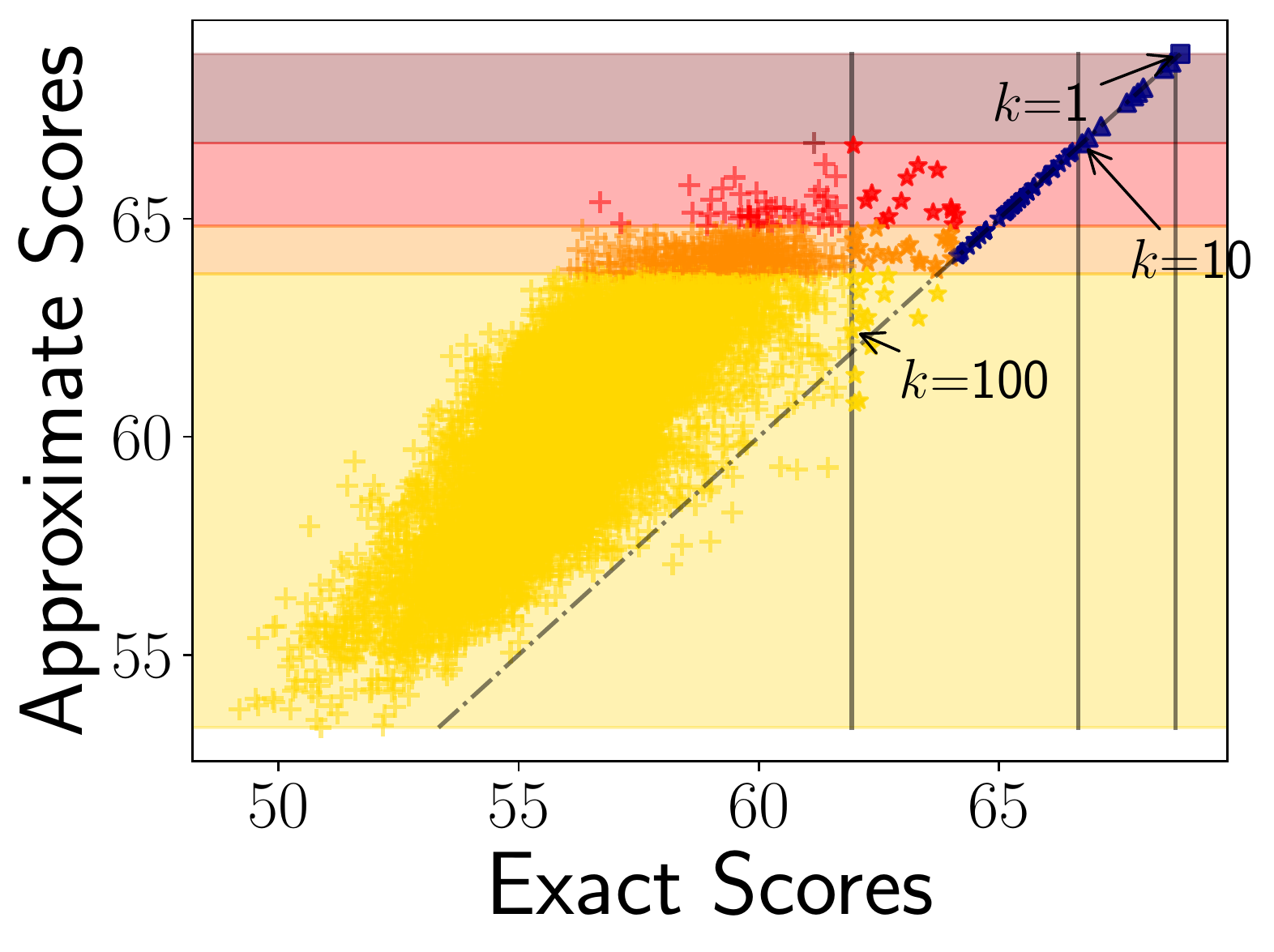}
        \caption{$\epsTopkGTSample{0}{\epsilon}$, $\epsilon=0.0$}  
        \label{fig:rq_6_example_exact_sampling_greedy_yugioh}
    \end{subfigure}
    \hfill
    \begin{subfigure}[b]{0.23\textwidth}
         \includegraphics[width=\textwidth]{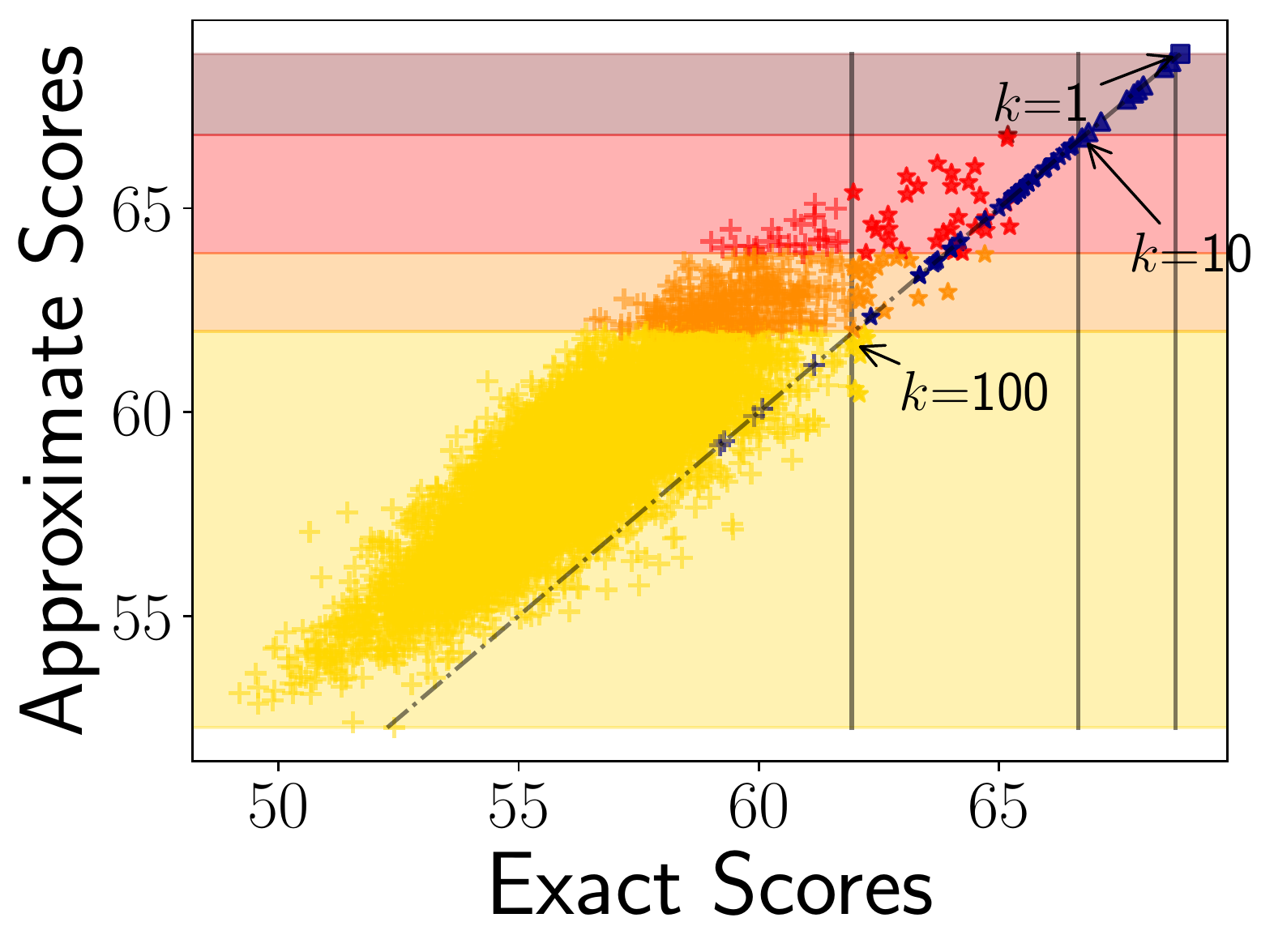}
        \caption{$\epsSoftmaxGTSample{0}{\epsilon}$, $\epsilon=0.0$}  
        \label{fig:rq_6_example_exact_sampling_smax_yugioh}
    \end{subfigure}
    \hfill 
    \begin{subfigure}[b]{0.23\textwidth}
         \includegraphics[width=\textwidth]{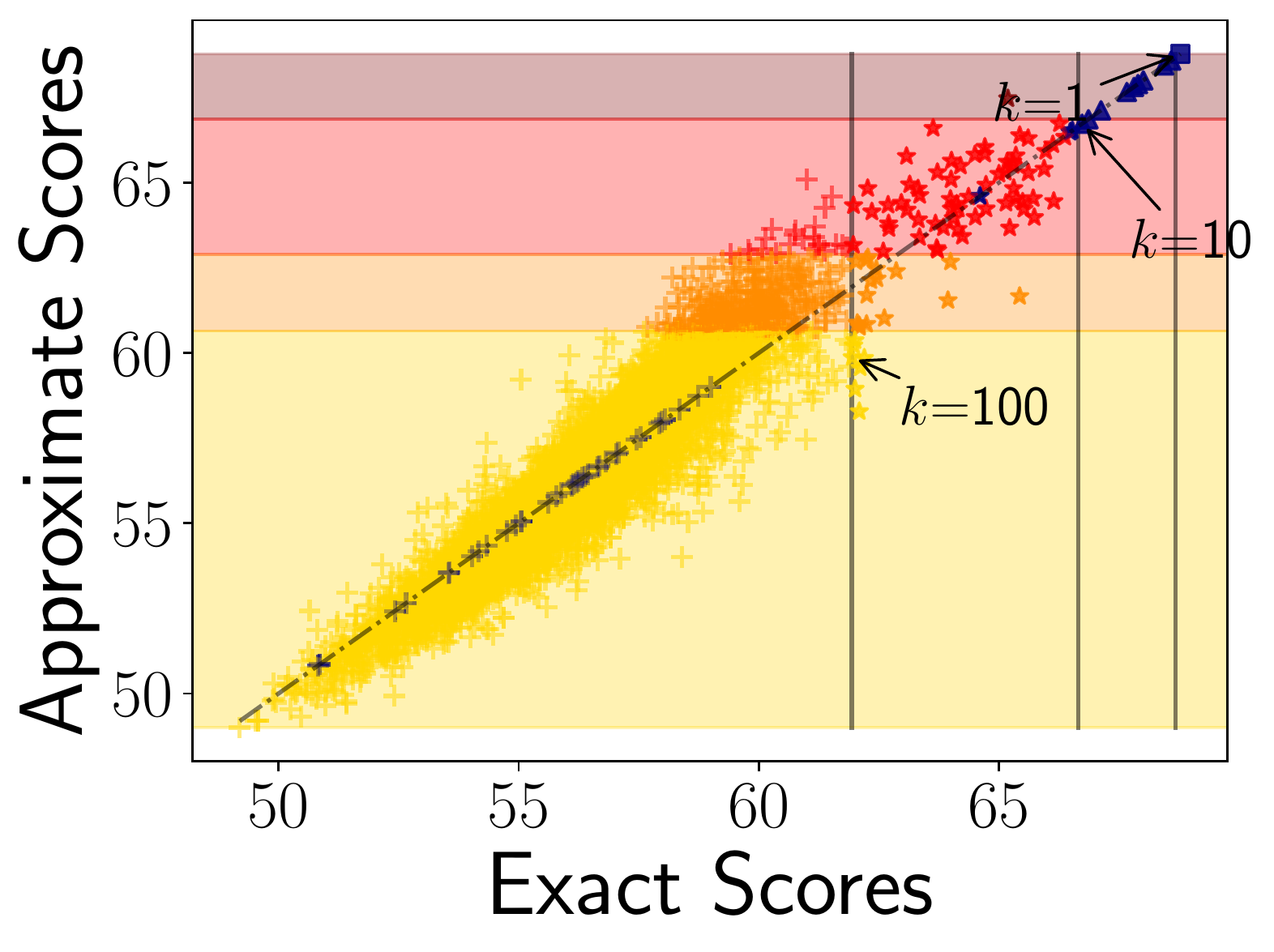}
        \caption{$\epsTopkGTSample{0}{\epsilon}$, $\epsilon=0.75$}
         \label{fig:rq_6_example_exact_sampling_greedy_rand_eps_75_yugioh}
    \end{subfigure}
    \hfill
    \begin{subfigure}[b]{0.23\textwidth}
         \includegraphics[width=\textwidth]{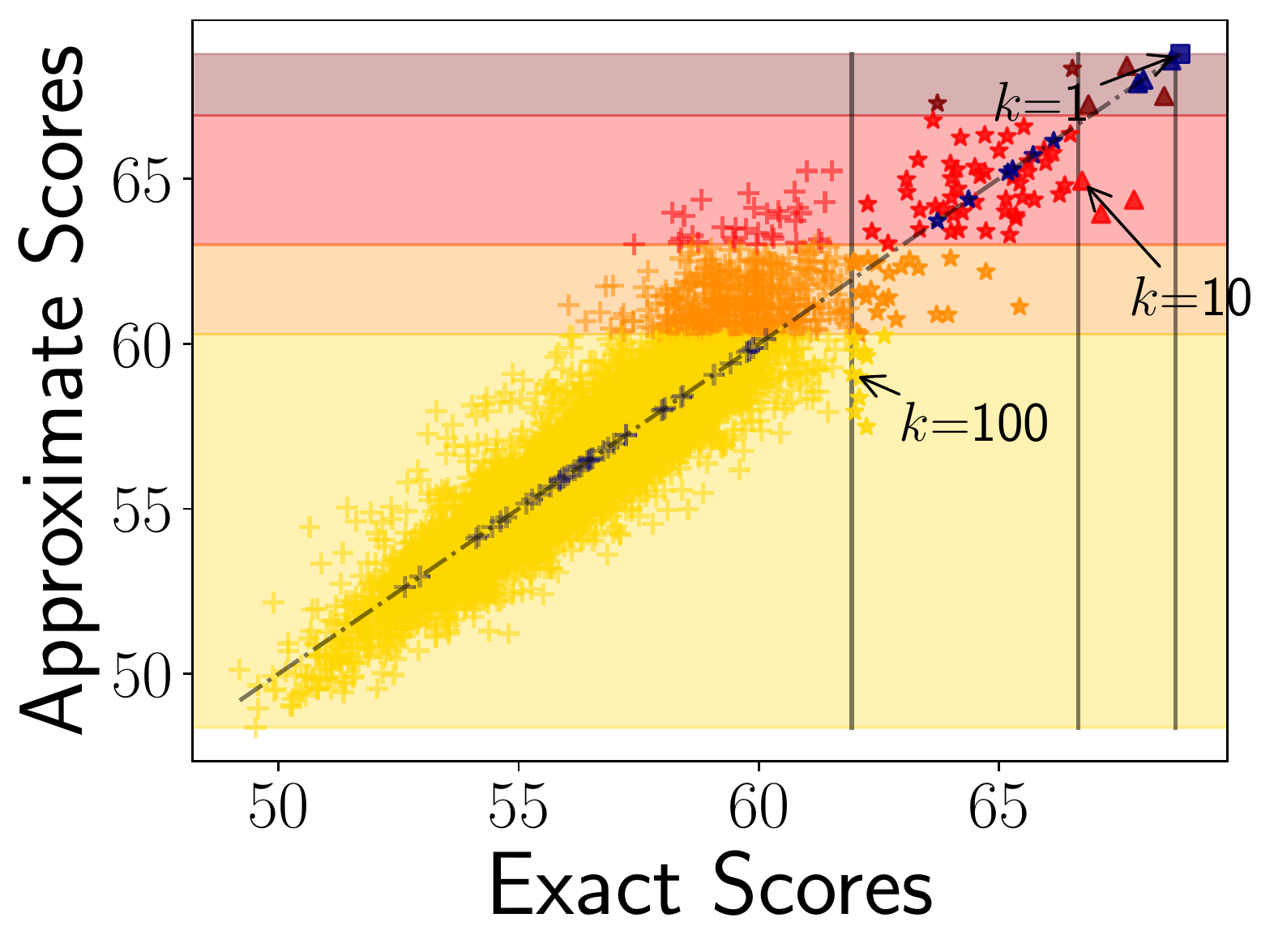}
        \caption{$\epsSoftmaxGTSample{0}{\epsilon}$,  $\epsilon=0.75$}  
        \label{fig:rq_6_example_exact_sampling_smax_rand_eps_75_yugioh}
    \end{subfigure}
    \caption{
    Scatter plot showing approximate versus exact cross-encoder scores 
    for a query from domain=\yugioh,
    when choosing $\nAnchorItems=50$ anchor items using oracle strategies from~\S\ref{apndx_subsec:oracle_exp} and $\queryTrainSize=500$. Top-$k$ for 
    $k$=1,10,100 wrt exact cross-encoder scores are annotated with text along with vertical lines, different
    color bands indicate the ordering of items wrt approximate scores,
    and anchor items are shown in blue.
    }
    \label{fig:rq_6_example_exact_sampling_method_yugioh}
    \vspace{-0.6cm}
\end{figure}

%% file: figs/appendix_rq_10_multi_vec.tex
\begin{figure}[!t]
    \centering
    \begin{subfigure}[b]{0.47\textwidth}
        \centering
        \includegraphics[width=\textwidth]{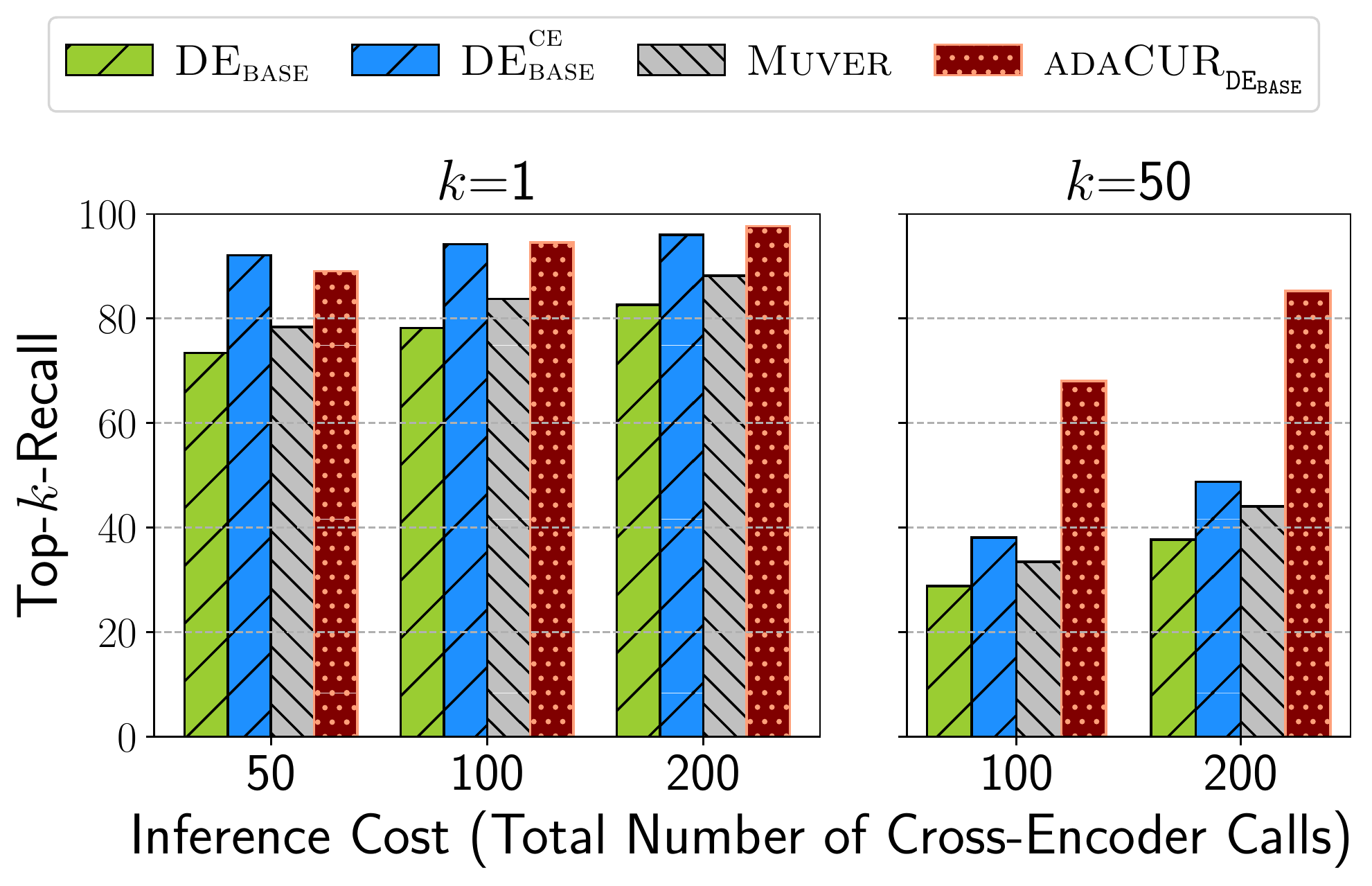}
        \caption{\yugioh}
        \label{apndx_fig:rq_10_multi_vec_yugioh}
    \end{subfigure}
    \begin{subfigure}[b]{0.47\textwidth}
        \centering
        \includegraphics[width=\textwidth, trim={0 0 0 2cm}, clip]{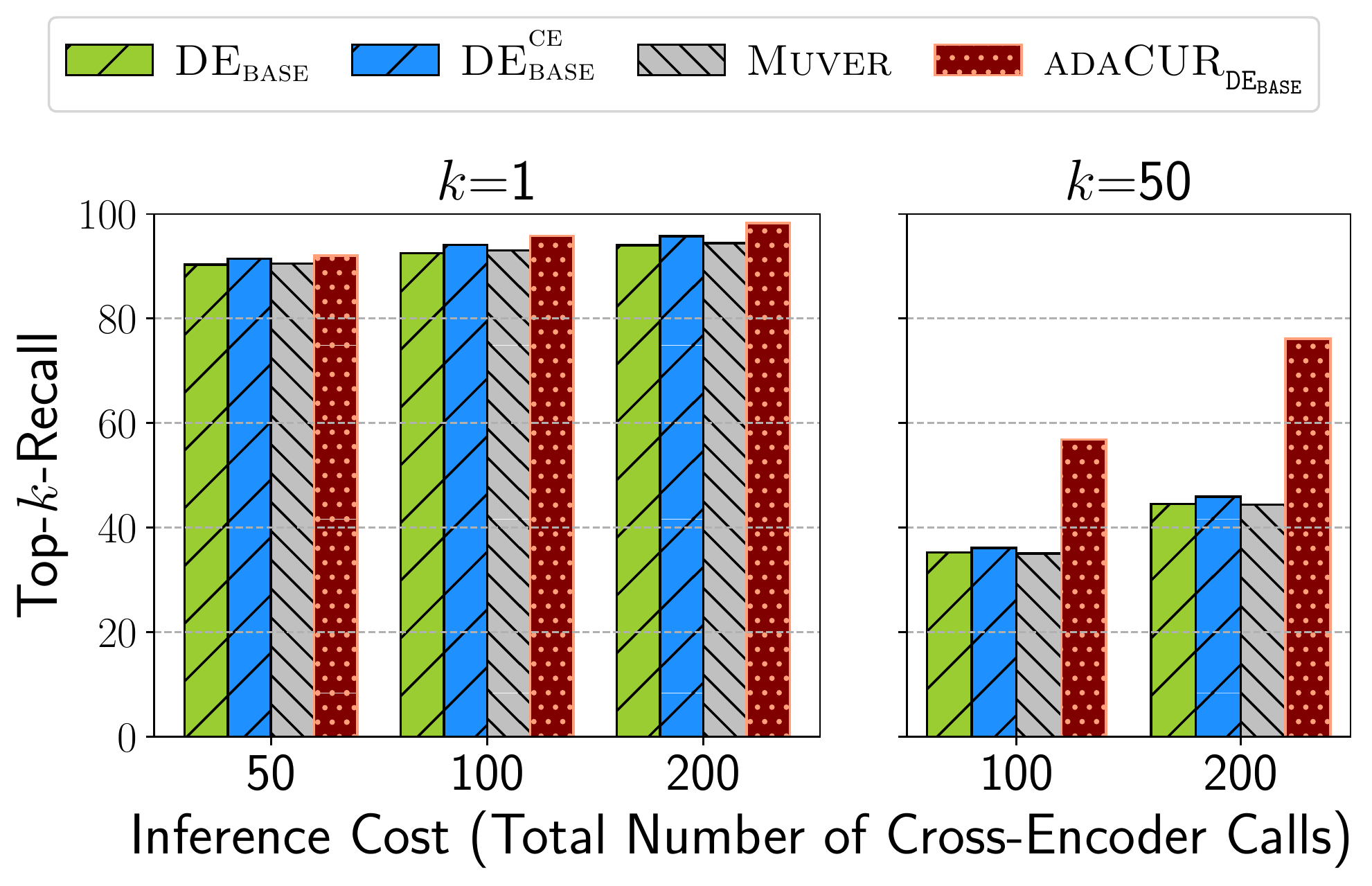}
        \caption{\starTrek}
        \label{apndx_fig:rq_10_multi_vec_star_trek}
    \end{subfigure}
    \caption{Top-$k$-Recall for \adaCURwFixedDE and baselines including multi-vector models on \zeshel domains, $\queryTrainSize=2000$. See \S\ref{apndx_subsec:multi_vec_results} for discussion.}
    \label{apndx_fig:rq_10_multi_vec_zeshel}
    \vspace{-0.5cm}
\end{figure}

%% file: figs/appendix_per_round_scatter.tex
\begin{figure*}[!ht]
    \centering
    \begin{subfigure}[b]{0.5\textwidth}
         \includegraphics[width=\textwidth]{figs/scatter_plots/legend.pdf}
        \phantomcaption{} 
    \end{subfigure}
    
    \begin{subfigure}[b]{\textwidth}
        \begin{subfigure}[b]{0.3\textwidth}
        \addtocounter{subfigure}{-1}
        \renewcommand\thesubfigure{\alph{subfigure}-1}
             \includegraphics[width=\textwidth]{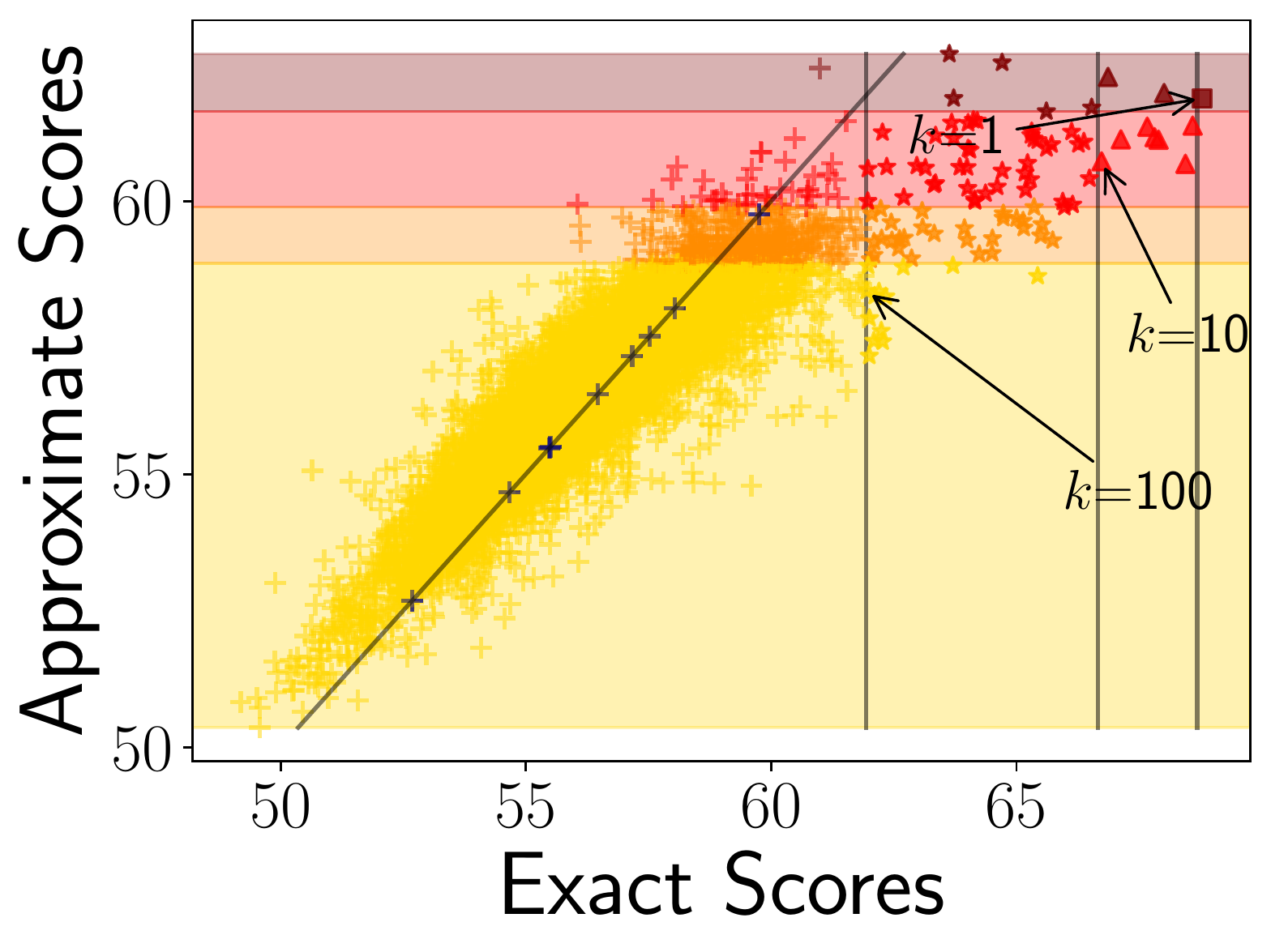}
            \caption{\adaCUR Round 1.\\ \centering{Total 10 anchor items.}}  
            \label{apndx_fig:example_icur_sampling_per_round_50_1}
        \end{subfigure}
        \hfill
        \begin{subfigure}[b]{0.3\textwidth}
            \addtocounter{subfigure}{-1}
            \renewcommand\thesubfigure{\alph{subfigure}-2}
             \includegraphics[width=\textwidth]{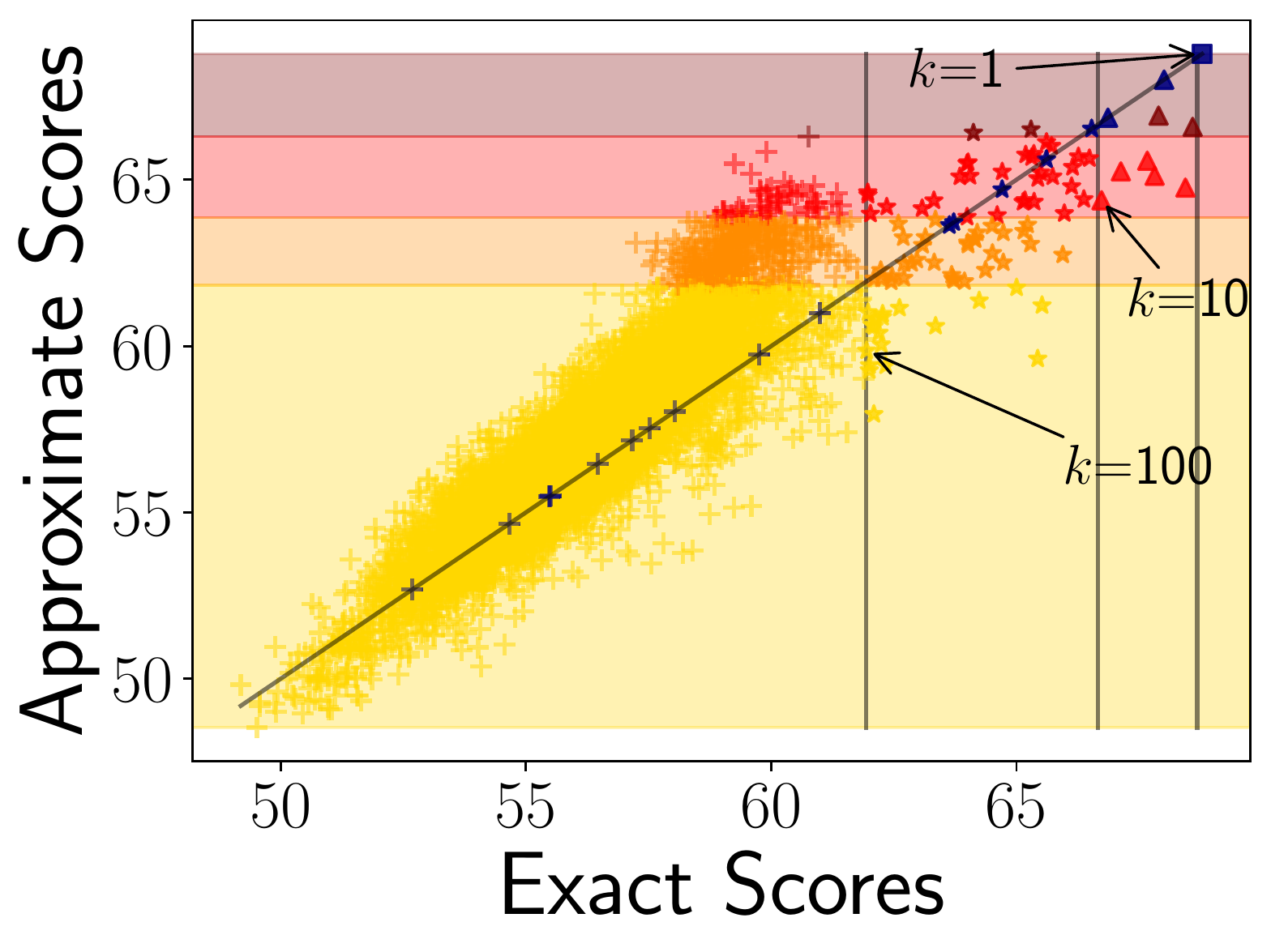}
            \caption{\adaCUR Round 2.\\ \centering{Total 20 anchor items.}}
            \label{apndx_fig:example_icur_sampling_per_round_50_2}
        \end{subfigure}
        \hfill
        \begin{subfigure}[b]{0.3\textwidth}
             \addtocounter{subfigure}{-1}
            \renewcommand\thesubfigure{\alph{subfigure}-3}
             \includegraphics[width=\textwidth]{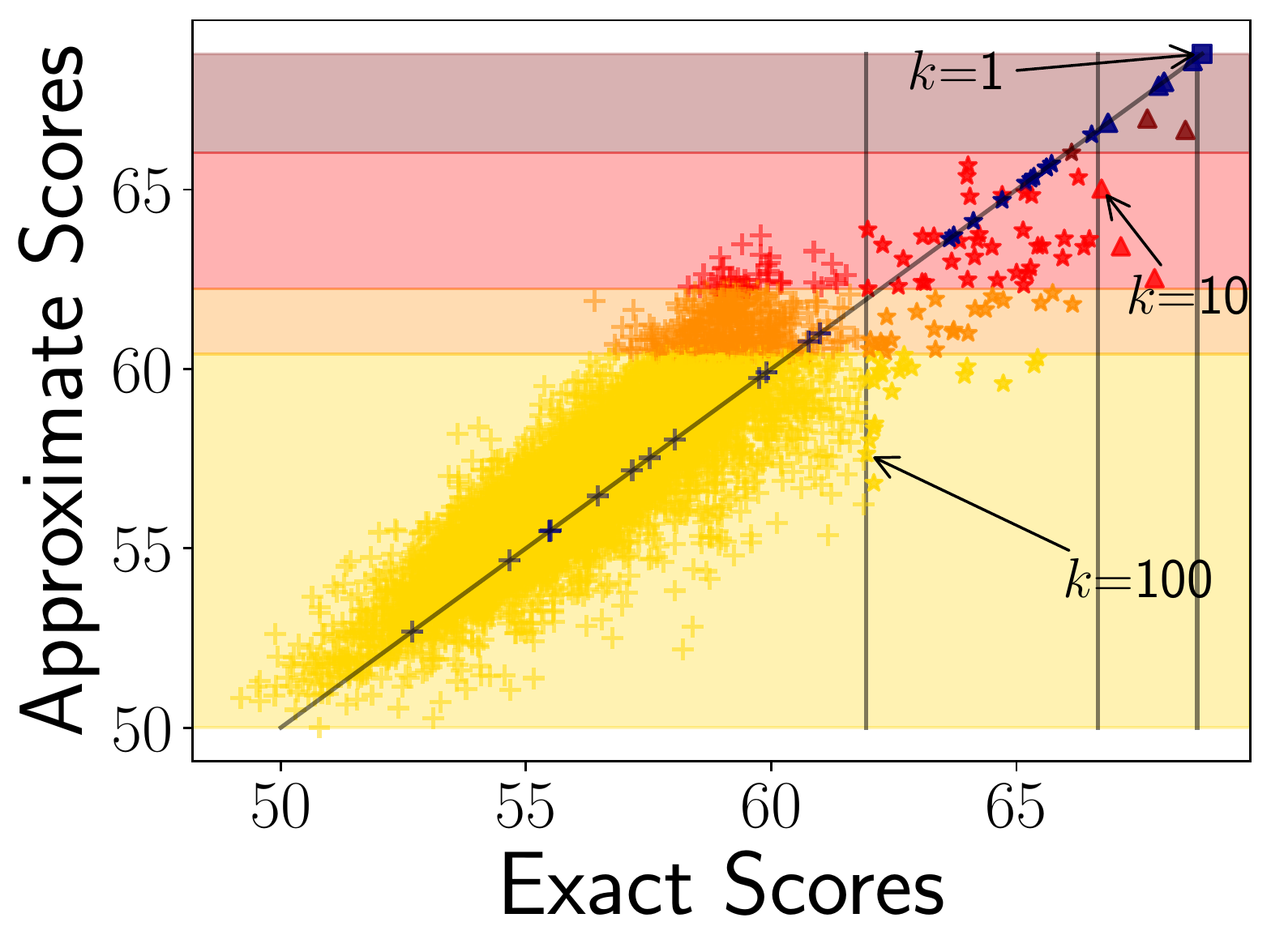}
            \caption{\adaCUR Round 3.\\ \centering{ Total 30 anchor items.}}
            \label{apndx_fig:example_icur_sampling_per_round_50_3}
        \end{subfigure}
        \\
        \begin{subfigure}[b]{0.3\textwidth}
            \addtocounter{subfigure}{-1}
            \renewcommand\thesubfigure{\alph{subfigure}-4}
             \includegraphics[width=\textwidth]{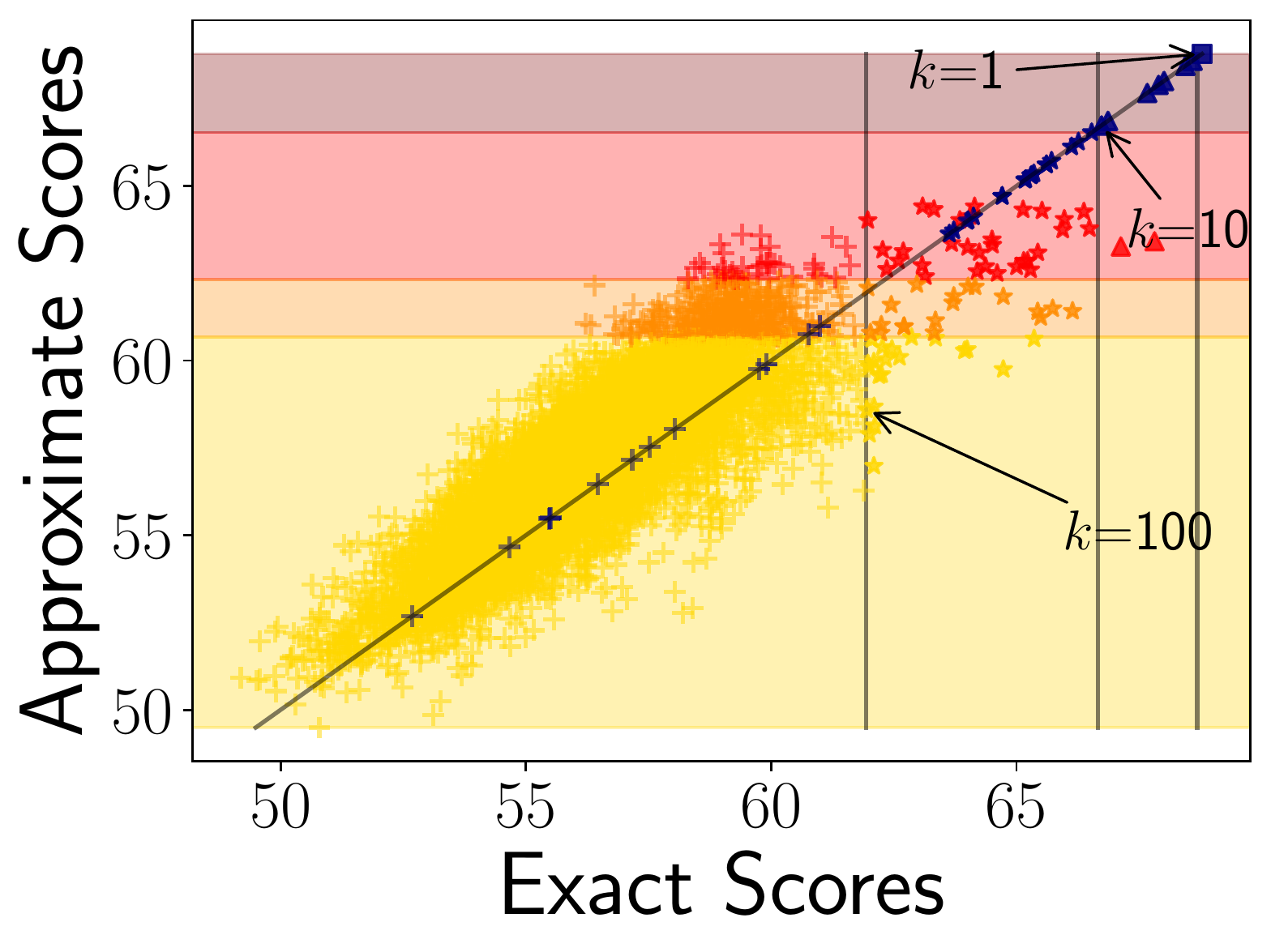}
            \caption{\adaCUR Round 4.\\ \centering{ Total 40 anchor items.}}
            \label{apndx_fig:example_icur_sampling_per_round_50_4}
        \end{subfigure}
        \hfill
        \begin{subfigure}[b]{0.3\textwidth}
            \addtocounter{subfigure}{-1}
            \renewcommand\thesubfigure{\alph{subfigure}-5}
             \includegraphics[width=\textwidth]{figs/appendix/scatter_plots/ment_idx=6_approx_topk_cumul_k_q_500_k_i_50_step=4.pdf}
            \caption{\adaCUR Round 5.\\ \centering{ Total 50 anchor items.}}
            \label{apndx_fig:example_icur_sampling_per_round_50_5}
        \end{subfigure}
        \hfill
        \begin{subfigure}[b]{0.3\textwidth}
            \addtocounter{subfigure}{-1}
            \renewcommand\thesubfigure{\alph{subfigure}-6}
             \includegraphics[width=\textwidth]{figs/scatter_plots/ment_idx_6/ment_idx=6_random_cumul_k_q_500_k_i_50.pdf}
            \caption{ \annCUR - Sampling all 50 anchor items uniformly at random.}
            \label{apndx_fig:example_anncur_sampling_50}
        \end{subfigure}
        \addtocounter{subfigure}{-1}
        \caption{Sampling 50 anchor items adaptively for \adaCUR (over five rounds) and for \annCUR (uniformly at random).}
    \end{subfigure}
    \\
    \begin{subfigure}[b]{\textwidth}
        \begin{subfigure}[b]{0.3\textwidth}
            \renewcommand\thesubfigure{\alph{subfigure}-1}
             \includegraphics[width=\textwidth]{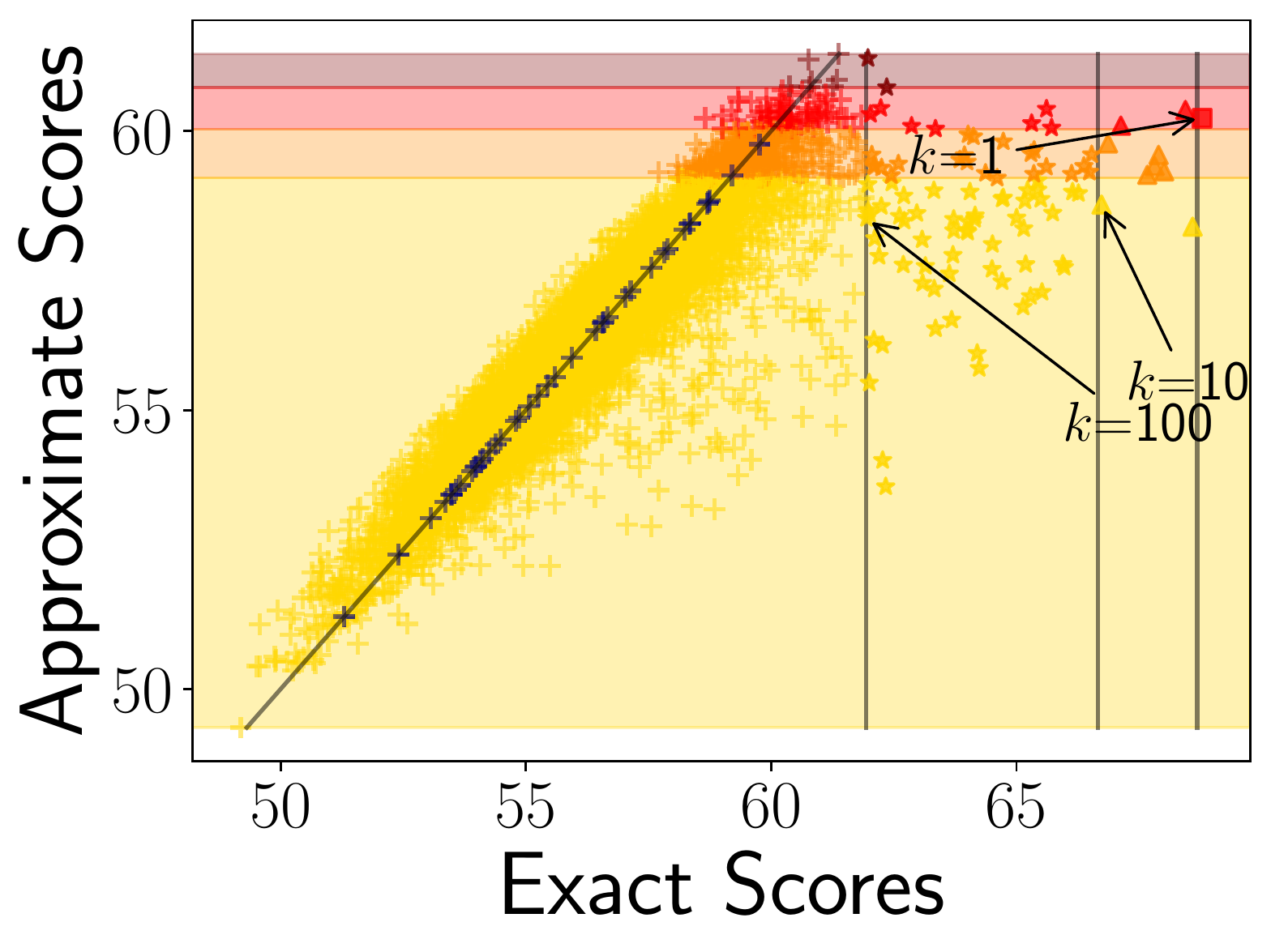}
            \caption{\adaCUR Round 1.\\ \centering{ Total 40 anchor items.}  }
            \label{apndx_fig:example_icur_sampling_per_round_200_1}
        \end{subfigure}
        \hfill
        \begin{subfigure}[b]{0.3\textwidth}
            \addtocounter{subfigure}{-1}
            \renewcommand\thesubfigure{\alph{subfigure}-2}
             \includegraphics[width=\textwidth]{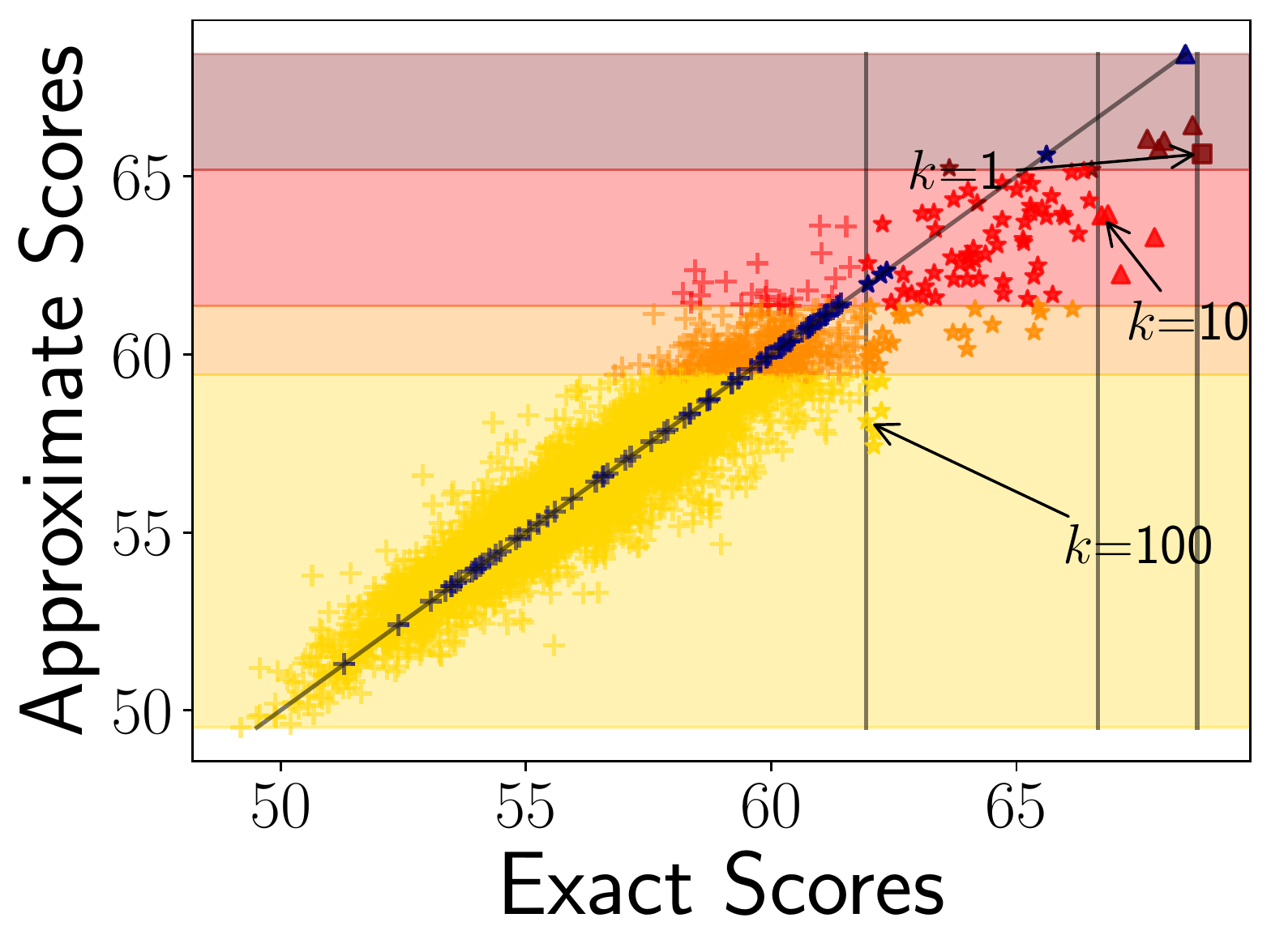}
            \caption{\adaCUR Round 2. \\ \centering{Total 80 anchor items.}}
            \label{apndx_fig:example_icur_sampling_per_round_200_2}
        \end{subfigure}
        \hfill
        \begin{subfigure}[b]{0.3\textwidth}
            \addtocounter{subfigure}{-1}
             \renewcommand\thesubfigure{\alph{subfigure}-3}
             \includegraphics[width=\textwidth]{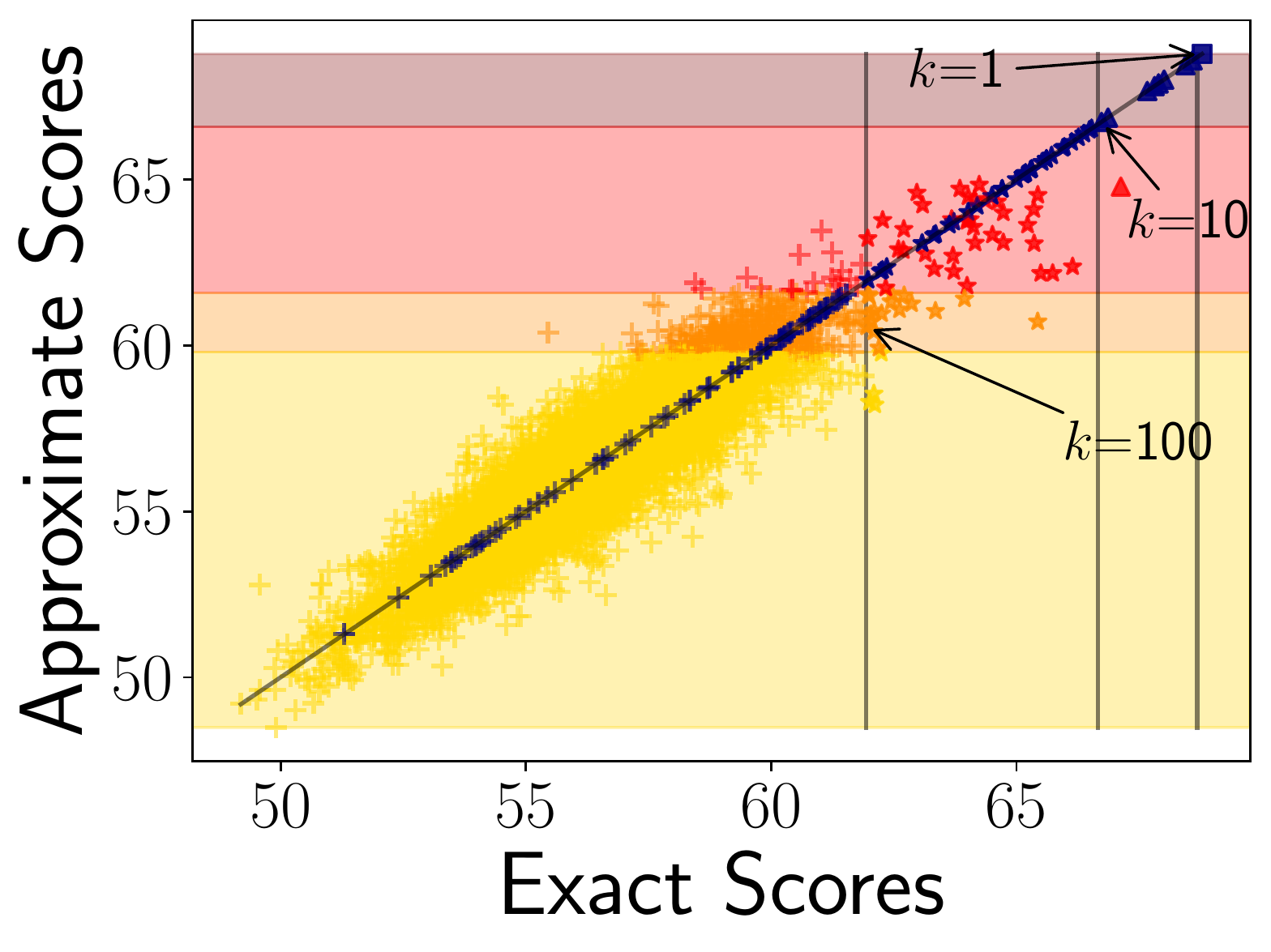}
            \caption{\adaCUR Round 3. \\ \centering{Total 120 anchor items.}}
            \label{apndx_fig:example_icur_sampling_per_round_200_3}
        \end{subfigure}
        \\
        \begin{subfigure}[b]{0.3\textwidth}
            \addtocounter{subfigure}{-1}
            \renewcommand\thesubfigure{\alph{subfigure}-4}
             \includegraphics[width=\textwidth]{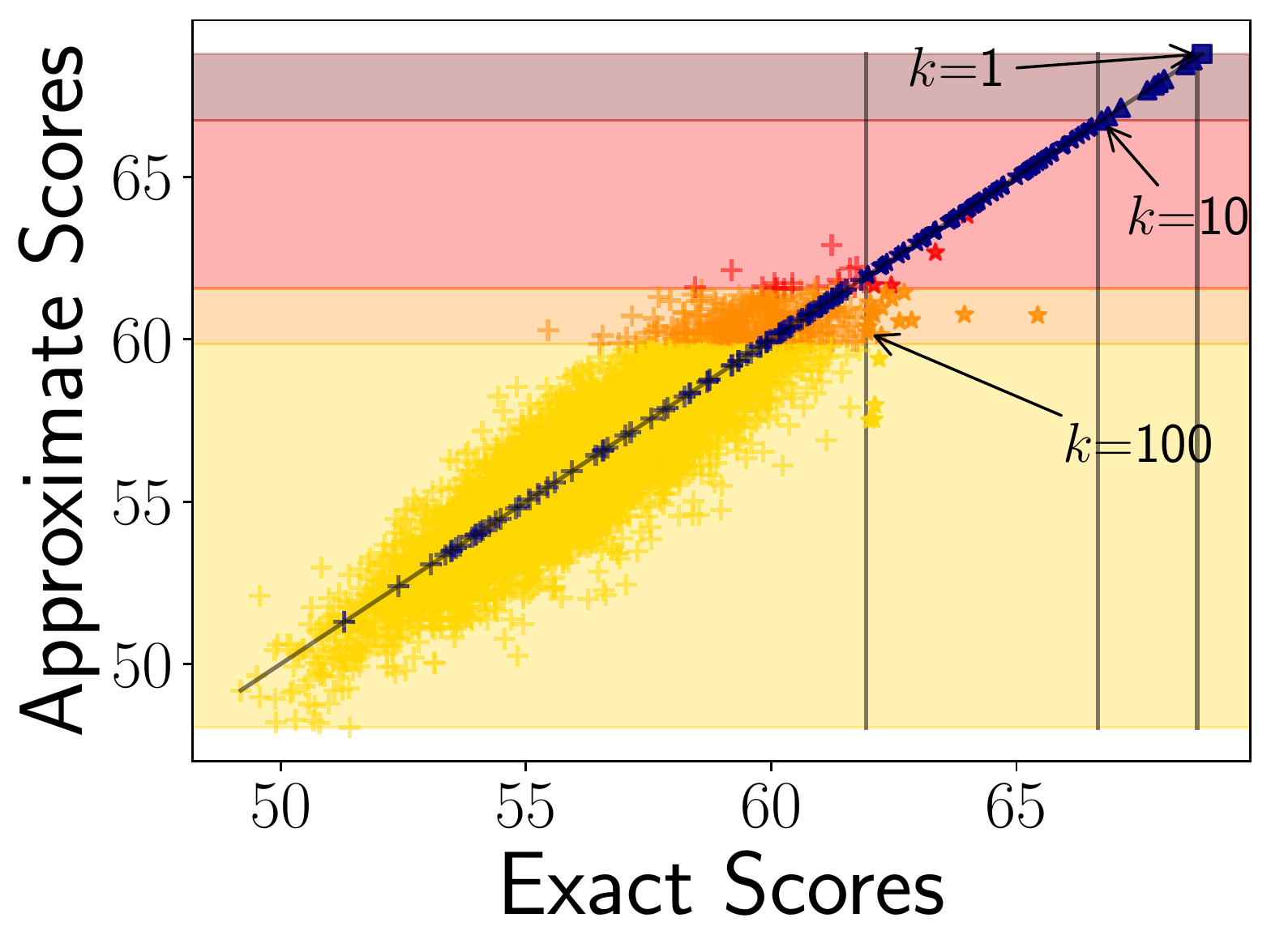}
            \caption{\adaCUR Round 4. \\ \centering{Total 160 anchor items.}}
            \label{apndx_fig:example_icur_sampling_per_round_200_4}
        \end{subfigure}
        \hfill
        \begin{subfigure}[b]{0.3\textwidth}
            \addtocounter{subfigure}{-1}
            \renewcommand\thesubfigure{\alph{subfigure}-5}

             \includegraphics[width=\textwidth]{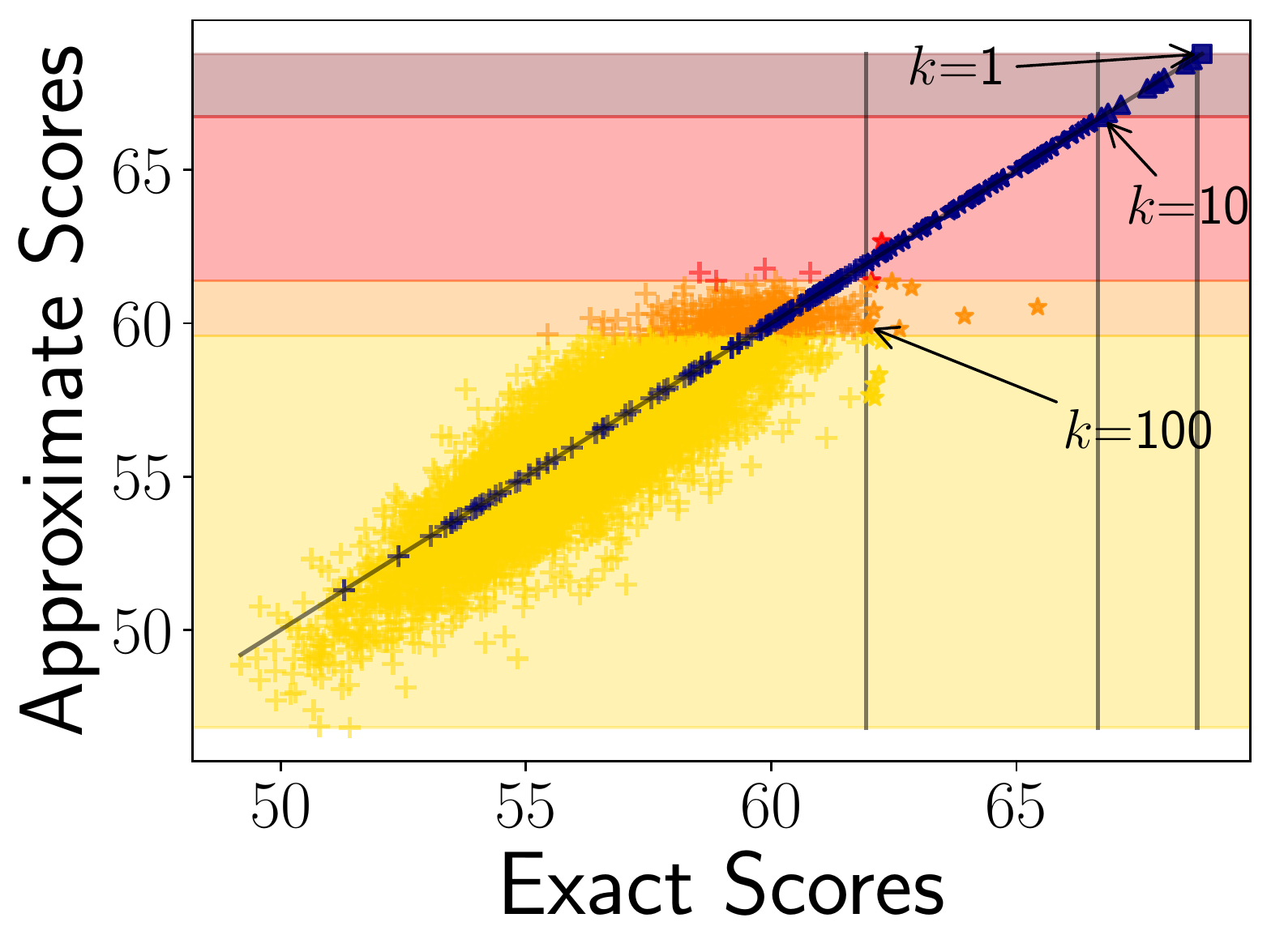}
            \caption{\adaCUR Round 5. \\ \centering{Total 200 anchor items.}}
            \label{apndx_fig:example_icur_sampling_per_round_200_5}
        \end{subfigure}
        \hfill
        \begin{subfigure}[b]{0.3\textwidth}
            \addtocounter{subfigure}{-1}
            \renewcommand\thesubfigure{\alph{subfigure}-6}
             \includegraphics[width=\textwidth]{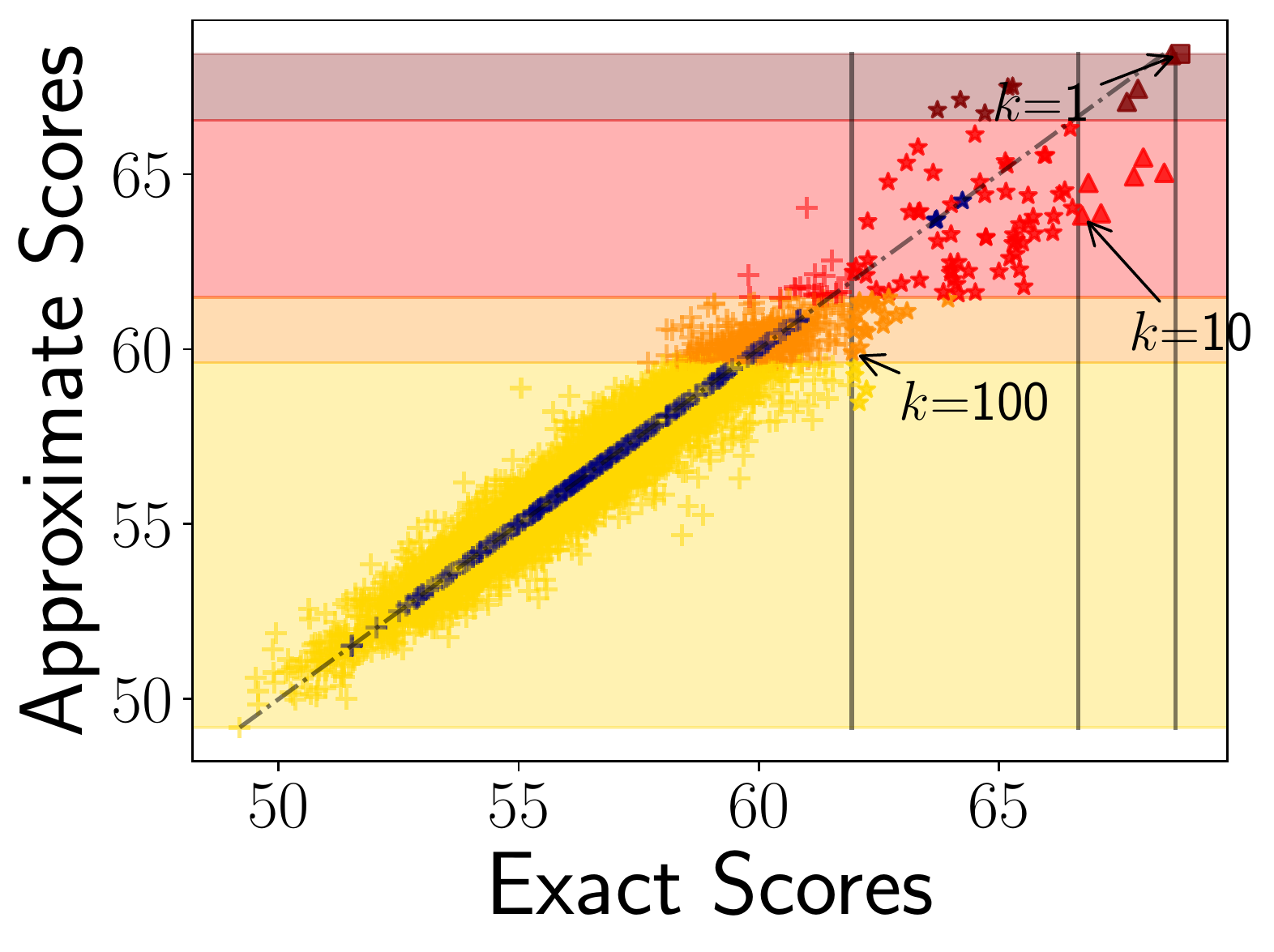}
            \caption{\annCUR - Sampling all 200 anchor items uniformly at random.}
            \label{apndx_fig:example_anncur_sampling_200}
        \end{subfigure}
        \addtocounter{subfigure}{-1}
        \caption{Sampling 200 anchor items adaptively for \adaCUR (over five rounds) and for \annCUR (uniformly at random).}
    \end{subfigure}
    \caption{
    Scatter plot showing approximate versus exact cross-encoder scores 
    for a query from domain=\yugioh,$\queryTrainSize=500$
    when choosing $\nAnchorItems=50$ and $200$ anchor items with \adaCUR over five rounds, and uniformly at random with \annCUR.
    Top-$k$ for $k$=1,10,100 wrt exact cross-encoder scores are annotated with text, different
    color bands indicate the ordering of items wrt approximate scores,
    and anchor items are shown in blue. 
    With \adaCUR, the first batch containing anchor items in 
    Figure~\ref{apndx_fig:example_icur_sampling_per_round_50_1} 
    and \ref{apndx_fig:example_icur_sampling_per_round_200_1} 
    is chosen uniformly at random and in subsequent rounds, items with highest approximate scores are chosen. 
    Note that the approximation error for top-scoring items improves significantly when the 50 anchor items are chosen adaptively (see Figure~\ref{apndx_fig:example_icur_sampling_per_round_50_5})  with the improvement being much more significant than merely increasing the number of anchor items sampled uniformly at random from 50 in Figure~\ref{apndx_fig:example_anncur_sampling_50} to 200 in Figure~\ref{apndx_fig:example_anncur_sampling_200}.
    }
    \label{apndx_fig:example_icur_sampling_per_round}
\end{figure*}

%% file: figs/appendix_rq_2_beir.tex
\begin{figure*}[!ht]

    \begin{subfigure}[b]{\textwidth}
        \includegraphics[width=\textwidth, trim={2cm 10.8cm 2cm 0}, clip]{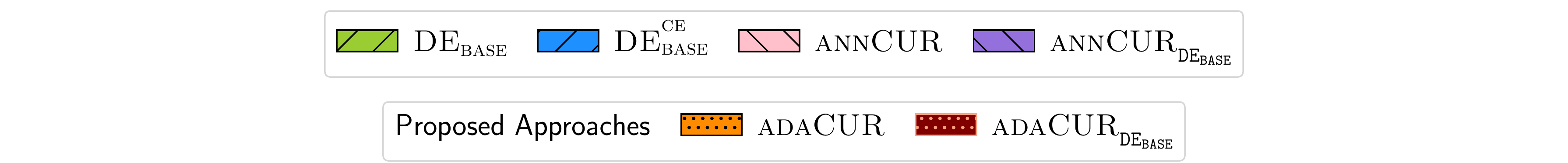}
    \end{subfigure}
    \vspace{0.5cm}
    \begin{subfigure}[b]{\textwidth}
        \includegraphics[width=\textwidth, trim={0 0 0 3.8cm}, clip]{figs/appendix/rq_2a_domain=scidocs_n_test_q=1000_n_trn_q=1000_n_steps=10_bienc=v2_recall_vs_cost_topk=1_10_100.pdf}
            
        \caption{\scidocs}
        \label{apndx_fig:rq_2_recall_at_same_cost_scidocs_1000}
    \end{subfigure}
    
    \begin{subfigure}[b]{\textwidth}
        \includegraphics[width=\textwidth, trim={0 0 0 3.8cm}, clip]{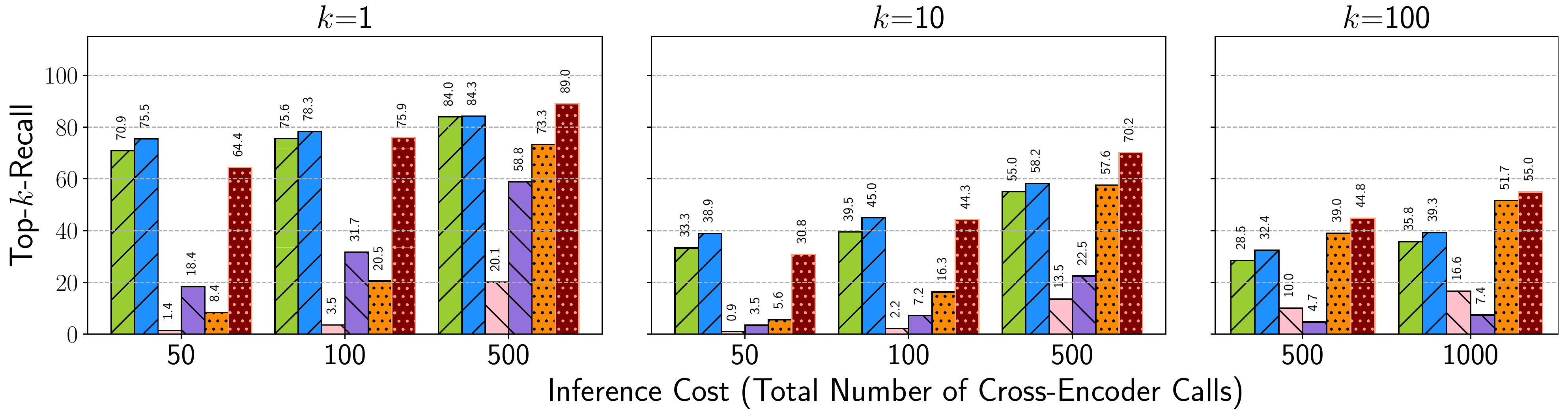}
            
        \caption{\hotpotqa}
        \label{apndx_fig:rq_2_recall_at_same_cost_hotpotqa_1000}
    \end{subfigure}
    \caption{Top-$k$-Recall for \adaCUR (using ten rounds) and baselines for \scidocs and \hotpotqa, $\queryTrainSize$=1000.}
    \label{apndx_fig:rq_2_recall_at_same_cost_beir}
    \vspace{-0.5cm}
\end{figure*}

%% file: figs/appendix_rq_2.tex
\begin{figure*}[!ht]
    \centering
    \begin{subfigure}[b]{\textwidth}
         \includegraphics[trim={0 10.7cm 0 0},clip,width=\textwidth]{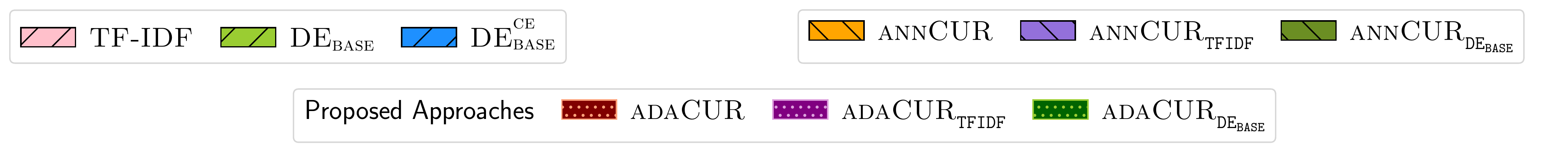}
        \phantomcaption{} 
    \end{subfigure}

    \addtocounter{subfigure}{-1}
    \begin{subfigure}[b]{\textwidth}
        \includegraphics[trim={0 0 0 3.7cm},clip,width=\textwidth]{figs/appendix/rq_2b_domain=yugioh_nm_train=100_n_steps=5_recall_vs_cost_topk=1_10_100.pdf}
        \caption{Number of train/anchor queries $\queryTrainSize=100$.}
        \label{apndx_fig:rq_2_recall_at_same_cost_yugioh_100}
    \end{subfigure}
    \\
    \vspace{0.5cm}
    \begin{subfigure}[b]{\textwidth}
        \includegraphics[trim={0 0 0 3.7cm},clip,width=\textwidth]{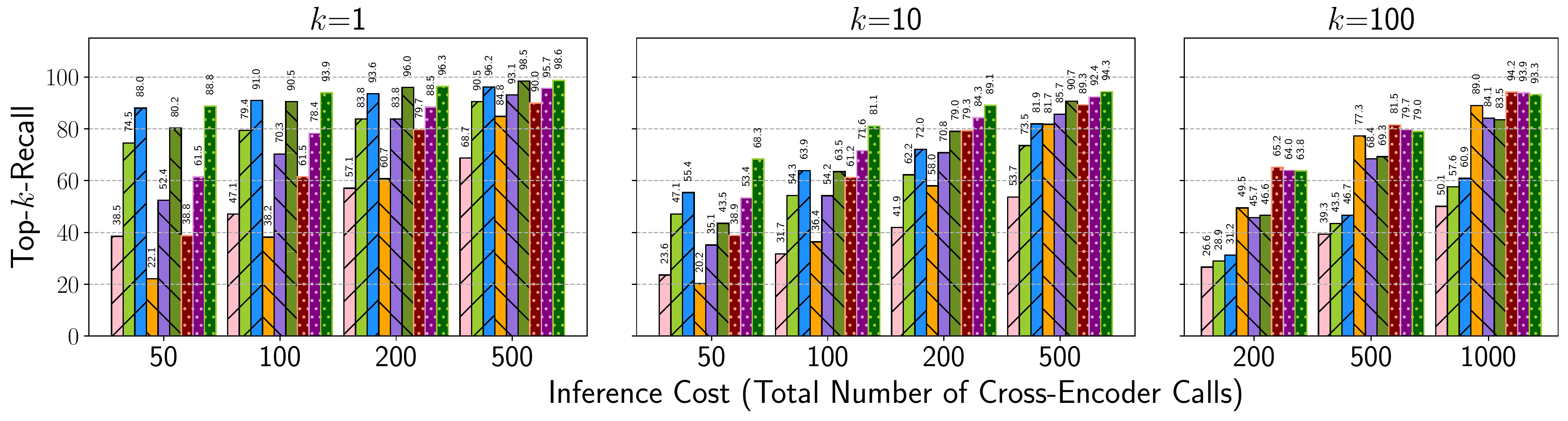}
        \caption{Number of train/anchor queries $\queryTrainSize=500$.}
        \label{apndx_fig:rq_2_recall_at_same_cost_yugioh_500}
    \end{subfigure}
    \\
    \vspace{0.5cm}
    \begin{subfigure}[b]{\textwidth}
        \includegraphics[trim={0 0 0 3.7cm},clip,width=\textwidth]{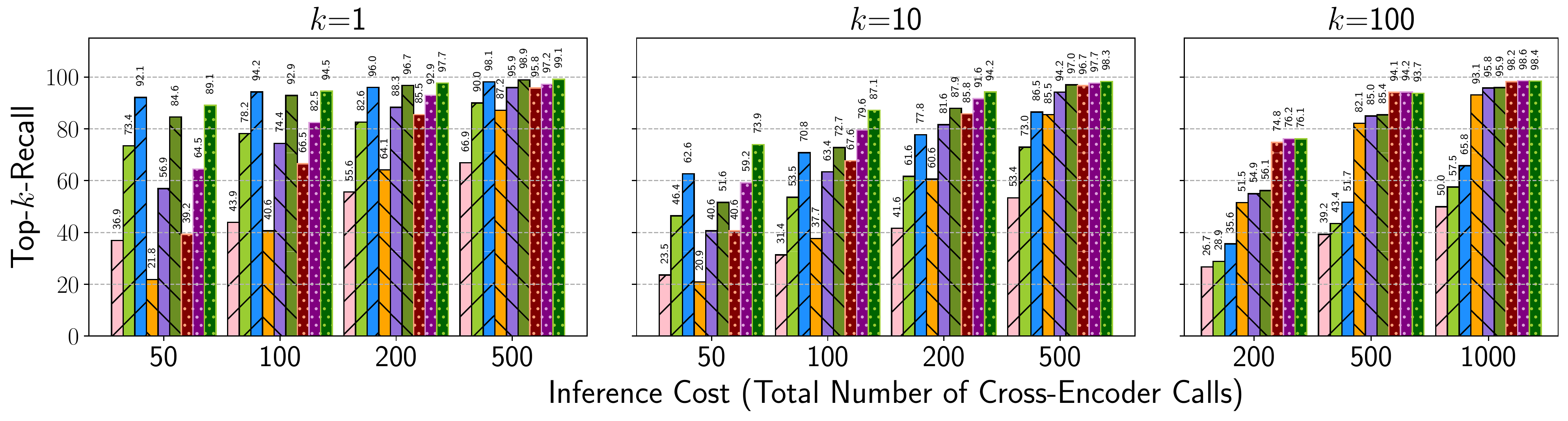}
        \caption{Number of train/anchor queries $\queryTrainSize=2000$.}
        \label{apndx_fig:rq_2_recall_at_same_cost_yugioh_2000}
    \end{subfigure}
    \caption{Top-$k$-Recall for \adaCUR (using five rounds) and baselines for domain=\yugioh. Each subfigure corresponds to a different value of the number of train/anchor queries ($\queryTrainSize$).}
    \label{apndx_fig:rq_2_recall_at_same_cost_yugioh}
\end{figure*}

\begin{figure*}[!ht]
    \centering
    \begin{subfigure}[b]{\textwidth}
         \includegraphics[trim={0 10.7cm 0 0},clip,width=\textwidth]{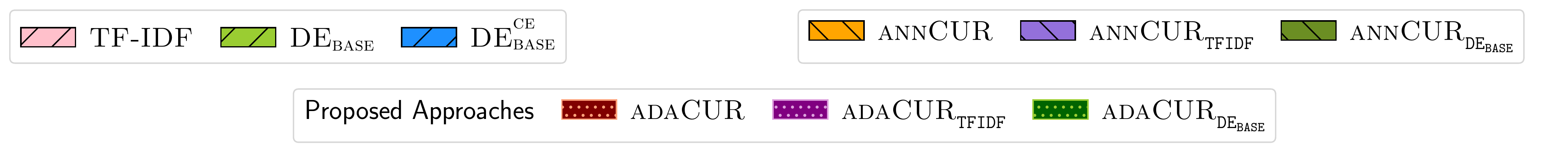}
        \phantomcaption{} 
    \end{subfigure}

    \addtocounter{subfigure}{-1}
    \begin{subfigure}[b]{\textwidth}
        \includegraphics[trim={0 0 0 3.7cm},clip,width=\textwidth]{figs/appendix/rq_2b_domain=star_trek_nm_train=100_n_steps=5_recall_vs_cost_topk=1_10_100.pdf}
        \caption{Number of train/anchor queries $\queryTrainSize=100$.}
        \label{apndx_fig:rq_2_recall_at_same_cost_starTrek_100}
    \end{subfigure}
    \\
    \vspace{0.5cm}
    \begin{subfigure}[b]{\textwidth}
        \includegraphics[trim={0 0 0 3.7cm},clip,width=\textwidth]{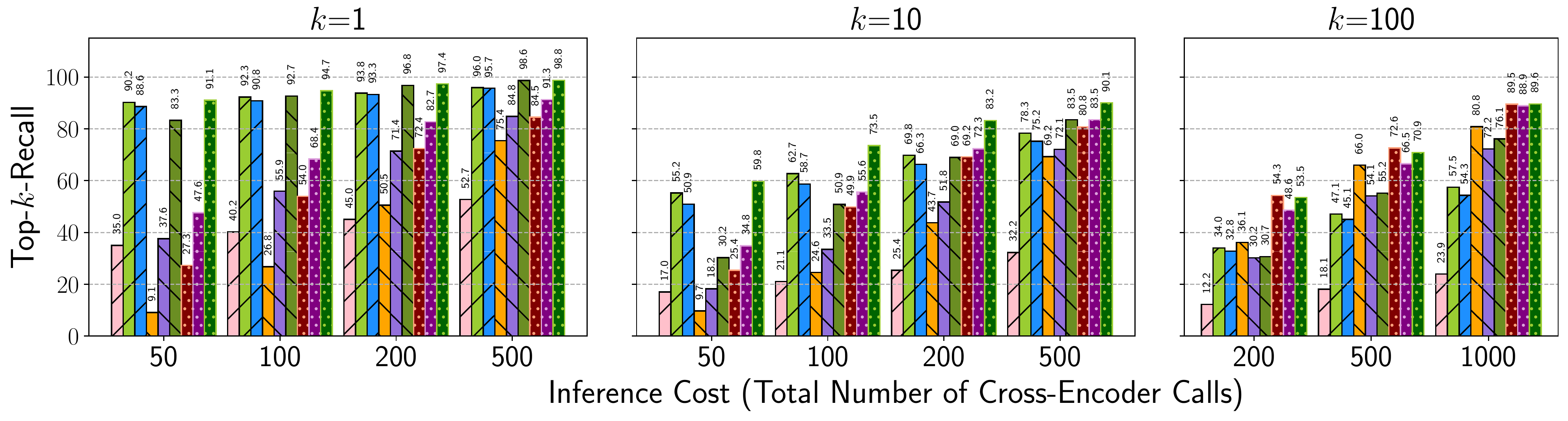}
        \caption{Number of train/anchor queries $\queryTrainSize=500$.}
        \label{apndx_fig:rq_2_recall_at_same_cost_starTrek_500}
    \end{subfigure}
    \\
    \vspace{0.5cm}
    \begin{subfigure}[b]{\textwidth}
        \includegraphics[trim={0 0 0 3.7cm},clip,width=\textwidth]{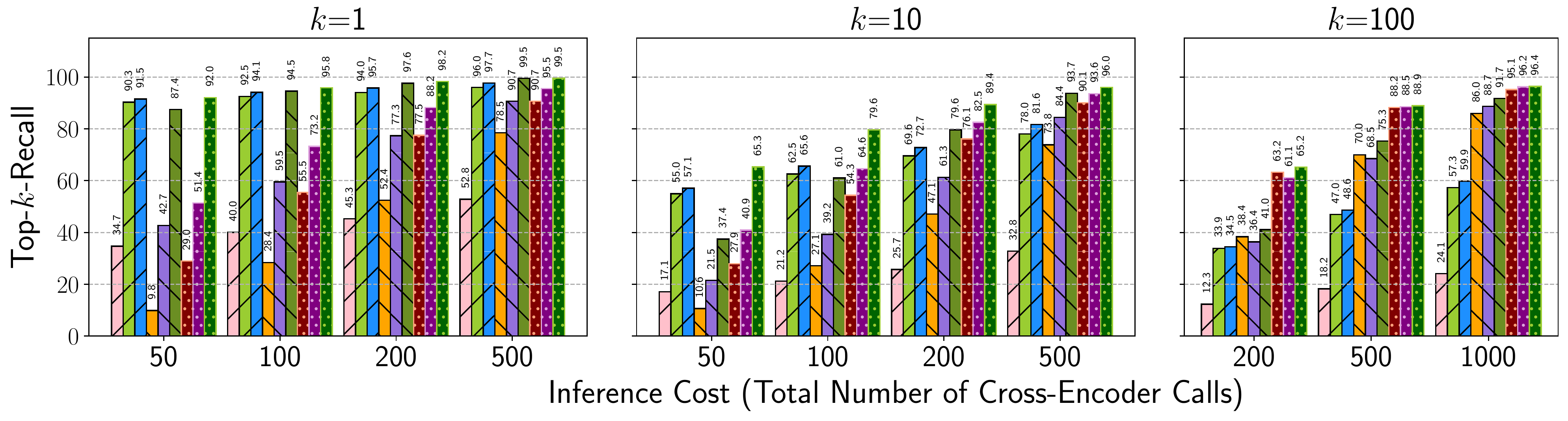}
        \caption{Number of train/anchor queries $\queryTrainSize=2000$.}
        \label{apndx_fig:rq_2_recall_at_same_cost_starTrek_2000}
    \end{subfigure}
    \caption{Top-$k$-Recall for \adaCUR (using five rounds) and baselines for domain=\starTrek. Each subfigure corresponds to a different value of the number of train/anchor queries ($\queryTrainSize$).}
    \label{apndx_fig:rq_2_recall_at_same_cost_starTrek}
\end{figure*}

\begin{figure*}[!ht]
    \centering
    \begin{subfigure}[b]{\textwidth}
         \includegraphics[trim={0 10.7cm 0 0},clip,width=\textwidth]{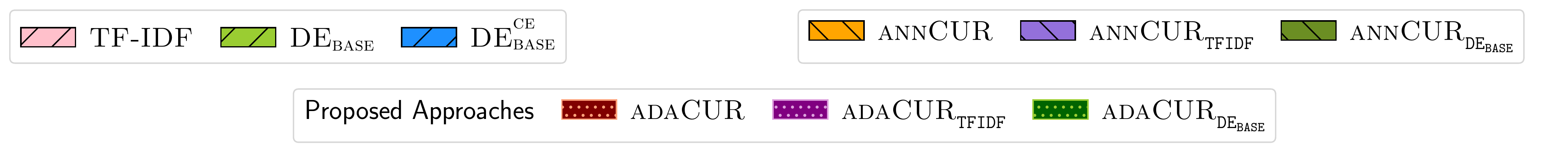}
        \phantomcaption{} 
    \end{subfigure}

    \addtocounter{subfigure}{-1}
    \begin{subfigure}[b]{\textwidth}
        \includegraphics[trim={0 0 0 3.7cm},clip,width=\textwidth]{figs/appendix/rq_2b_domain=military_nm_train=100_n_steps=5_recall_vs_cost_topk=1_10_100.pdf}
        \caption{Number of train/anchor queries $\queryTrainSize=100$.}
        \label{apndx_fig:rq_2_recall_at_same_cost_military_100}
    \end{subfigure}
    \\
    \vspace{0.5cm}
    \begin{subfigure}[b]{\textwidth}
        \includegraphics[trim={0 0 0 3.7cm},clip,width=\textwidth]{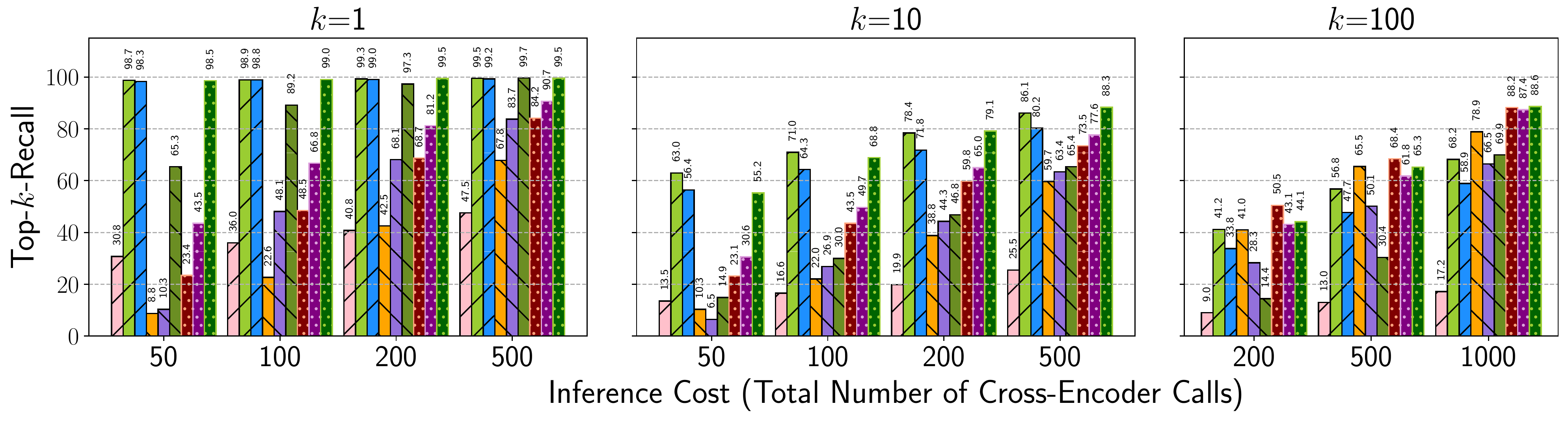}
        \caption{Number of train/anchor queries $\queryTrainSize=500$.}
        \label{apndx_fig:rq_2_recall_at_same_cost_military_500}
    \end{subfigure}
    \\
    \vspace{0.5cm}
    \begin{subfigure}[b]{\textwidth}
        \includegraphics[trim={0 0 0 3.7cm},clip,width=\textwidth]{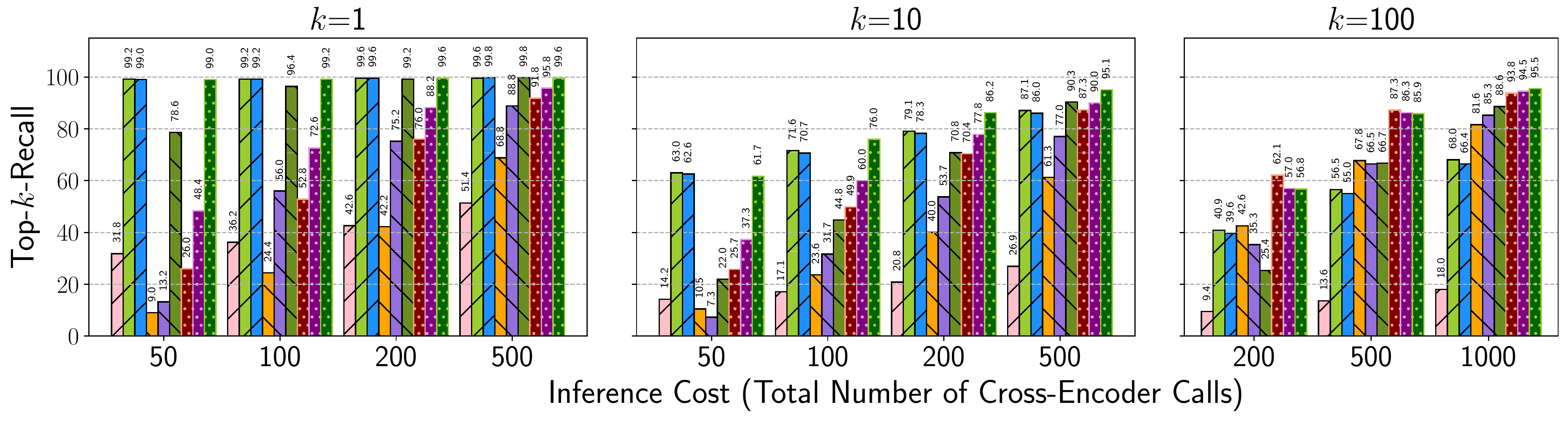}
        \caption{Number of train/anchor queries $\queryTrainSize=2000$.}
        \label{apndx_fig:rq_2_recall_at_same_cost_military_2000}
    \end{subfigure}
    \caption{Top-$k$-Recall for \adaCUR (using five rounds) and baselines for domain=\military. Each subfigure corresponds to a different value of the number of train/anchor queries ($\queryTrainSize$).
    Note that \fixedDualEncoder has high Top-1-Recall values as
    domain=\military is included in the set of train domains in \zeshel which
    are used to train both \fixedDualEncoder and the cross-encoder model.
    }
    \label{apndx_fig:rq_2_recall_at_same_cost_military}
\end{figure*}